\documentclass[a4paper,11pt]{article}
\usepackage{amssymb,amsmath,amsthm}
\usepackage{textcomp}

\usepackage{color}
\usepackage{latexsym}
\usepackage[english]{babel} 
\usepackage[latin1]{inputenc}  
\usepackage[T1]{fontenc}   
\usepackage[all]{xy}
\usepackage{epstopdf}
\usepackage{authblk}
\usepackage{stmaryrd}
\usepackage{amsfonts}
\usepackage{graphics}
\usepackage{epsfig}
\usepackage{graphicx}
\usepackage{epstopdf}
\usepackage{braket}
\usepackage{adjustbox}
\usepackage{tikz}
\usetikzlibrary{shapes,snakes}
\usepackage{amsmath,amssymb}
\usepackage{verbatim}
\newcommand{\du}{\node[rectangle] {$\downarrow\uparrow$}}
\newcommand{\de}{\node[rectangle] {$\downarrow$\textcolor{white}{$\uparrow$}}}
\newcommand{\ue}{\node[rectangle] {$\uparrow$\textcolor{white}{$\uparrow$}}}
\newcommand{\ee}{\node[rectangle] {\textcolor{white}{$\uparrow\downarrow$}}}

\def\CC{\mathbb{C}}

\def\C{\mathbb{C}}
\def\ZZ{\mathbb{F}}
\def\Z{\mathbb{F}}

\def\PP{\mathbb{P}}
\def\SS{\mathbb{S}}

\def\F{\mathbb{F}}

\def\du{\node[rectangle] {$\downarrow\uparrow$};}

\def\de{\node[rectangle] {$\downarrow\color{white} \uparrow$};}
\def\ue{\node[rectangle] {$\uparrow\color{white}\downarrow$};}

\usepackage{tikz}
\usetikzlibrary{automata}
\usetikzlibrary{shadows}
\usetikzlibrary{arrows}
\usetikzlibrary{shapes}
\usetikzlibrary{fit}
\usetikzlibrary{matrix}
\newcommand{\incl}{\ar@{^{}-}}
\newcommand{\inclu}{\ar@{^{}.}}

\def\SL{\text{SL}}
\def\SU{\text{SU}}
\def\Sep{\text{Sep}}
\def\SLOCC{\text{SLOCC}}

\def\up{\uparrow}
\def\down{\downarrow}

\newtheorem{lemma}{Lemma       }
\newtheorem{definition}{Definition  }[section]
\newtheorem{theorem}{Theorem}

\newtheorem{ex}{Example     }[section]
\newtheorem{remark}{Remark}[section]

\title{Geometric constructions over $\CC$ and $\ZZ_2$ for Quantum Information}
\author{Fr\'ed\'eric Holweck \\ Laboratoire Interdisciplinaire Carnot de Bourgogne\\ ICB/UTBM UMR 6303 CNRS, University Bourgogne-Franche-Comt\'e}
\begin{document}

\maketitle

\begin{abstract}
In this review paper I present two geometric constructions of distinguished nature, one is over the field of complex numbers 
$\CC$ and 
the other one is over the two elements field $\F_2$. Both constructions  have been 
employed in the past fifteen years to describe two quantum paradoxes or two resources of quantum information: 
entanglement of pure multipartite systems on one side and contextuality on the other. 
Both geometric constructions are linked to representation of semi-simple Lie groups/algebras. To emphasize this aspect one explains on one hand how well-known results in 
representation theory allows one
to see all the classification of entanglement classes of various tripartite quantum systems  ($3$ qubits, $3$ fermions, $3$ bosonic qubits...) in a unified picture. On the 
other hand, one also shows how some 
weight diagrams of simple Lie groups are encapsulated 
in the geometry which deals with the commutation relations of the generalized $N$-Pauli group.
\end{abstract}

\tableofcontents
\section*{Introduction}

The aim of this paper is to provide an elementary introduction 
to a series of papers involving geometrical descriptions 
of  two different problems in quantum information theory: the classification of entanglement classes 
for pure multipartite quantum systems on one hand \cite{HLT, HLT2,HLT3,HL,LH,HLP,HJ} and the observable-based proofs of 
the Kochen-Specker Theorem on the other hand \cite{HS,PSH,HSL,LHS}. Apparently, these two problems have no direct connections to each other
and the geometrical constructions to describe them are of distinguished nature. We will use projective complex geometry to describe entanglement classes and we will work with 
finite geometry 
over the two elements field $\mathbb{F}_2$ to describe operator-based proofs of the Kochen-Specker Theorem. However, when we look at both geometries from a representation theory point 
of view, one observes 
that the same semi-simple Lie groups are acting behind the scene. 
This observation may invite us to look for a more direct (physical) connection between those two questions. 
In this presentation I will also try to give many references on related works. 
However, this will not be an exhaustive review on all possible links between  geometry and quantum information  and there will be 
some references missing.

Before 
going into the details of the geometry, let us recall how those two questions are  historically related to the question of the existence of Hidden Variables Theory.

In the history of the development of quantum science, the paradoxes raised by questioning the 
foundations of quantum physics turn out to  be considered 
as quantum resources  once they have been tested experimentally. The most famous example of such a change of status for a scientific question is, of course, 
the EPR paradox which started by a criticism of the foundation of quantum physics by Einstein Podolsky and Rosen \cite{EPR}. 

The famous EPR paradox deals with what we nowadays call a pure $2$-qubit quantum system. This is a physical system made of two parts $A$ and $B$ such that 
each part, or each particle, is a two-level quantum system. Mathematically, a pure $2$-qubit state is a vector of $\mathcal{H}=\CC^2_A\otimes \CC^2_B$.
Denote by $(\ket{0},\ket{1})$ the standard basis of the vector spaces
$\CC^2_A$ and $\CC^2_B$ and let $(\ket{00},\ket{01},\ket{10},\ket{11})$ be the 
associated basis of $\mathcal{H}_{AB}$. The laws of quantum mechanics tell us that $\ket{\psi}\in  \mathcal{H}_{AB}$ can be described as 
\begin{equation}\label{eq2qubit}
 \ket{\psi}_{AB}=a_{00}\ket{00}+a_{10}\ket{10}+a_{01}\ket{01}+a_{11}\ket{11}, 
\end{equation}
 with  $a_{ij}\in \CC$ and  $|a_{00}|^2+|a_{10}|^2+|a_{01}|^2+|a_{11}|^2=1$. Einstein, Podolsky and Rosen introduced the following admissible state 
 \begin{equation} \ket{EPR}=\dfrac{1}{\sqrt{2}}(\ket{00}+\ket{11}),\end{equation}
to argue that quantum mechanics was incomplete. The EPR reasoning consists of saying that, according to quantum 
mechanics, 
a measurement of particle $A$ will project the system $\ket{EPR}$ to either $\ket{00}$ or $\ket{11}$, fixing instantaneously the possible outcomes of the measurement of particle $B$ no matter how far the distance between particles $A$ and $B$ is.
This was characterized in \cite{EPR} as {\em spooky action at the distance} and according to Einstein, Podolsky and Rosen this was showing 
that {\em  hidden variables} were necessary to make the theory complete. Note that none of all $2$-qubit quantum states can produce 
a spooky action at the distance. If $\ket{\psi}_{AB}=(\alpha_A \ket{0}+\beta_A \ket{1})\otimes (\alpha_B\ket{0}+\beta_B\ket{1})$, then the measurement of particle $A$ has no impact on the state of particle $B$.
From Eq (\ref{eq2qubit}) one sees that the possibility to factorize a state $\ket{\psi}_{AB}$ translates to 
\begin{equation}
 a_{00}a_{11}-a_{01}a_{10}=0.
\end{equation}

This homogeneous equation defines a quadratic hypersurface in $\PP^3=\PP(\CC^2\otimes \CC^2)$, corresponding to the projectivization 
of the states that can be factorized; those states are called {\em non-entangled states}.
The complement of the quadric is the set of non-factorizable states, i.e. {\em entangled states}.
\begin{figure}[!h]
 \begin{center}
\includegraphics[width=10cm]{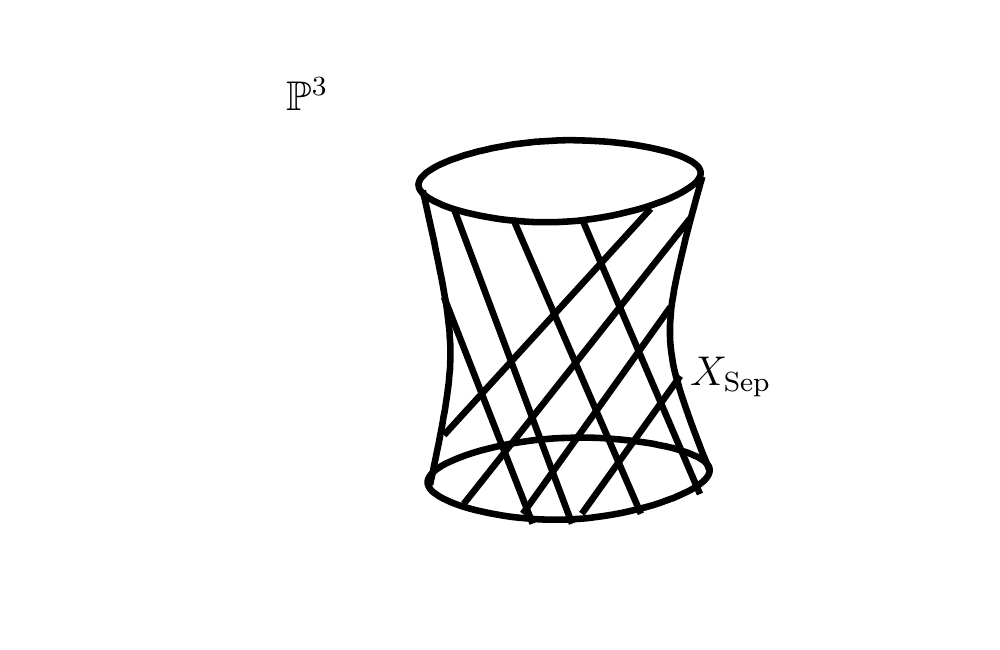}
 \caption{Non-entangled states, denoted by $X_\Sep$, and entangled states, $\PP^3\backslash X_\Sep$, in $\PP(\CC^2\otimes\CC^2)$.}
\end{center}
 \end{figure}

The philosophical questioning of Einstein and his 
co-authors about the existence of hidden-variables to make quantum physics complete 
becomes a scientific question after the work of John Bell \cite{bell}, sixty years later, whose inequalities have opened up the path to experimental tests. Those experimental tests have been performed many times starting 
with the pioneering works of Alain Aspect\cite{aspect} and {\em entanglement} in multipartite systems is nowadays recognized as an essential resource in quantum information.

Another paradox of quantum physics, maybe less famous than EPR, is {\em contextuality}. Interestingly, the notion of contextuality in quantum physics is also related to the question of
the existence of hidden-variables. In 1975 Kochen and Specker\footnote{This concept of contextuality also appears in Bell's paper \cite{bell,mermin}.} \cite{KS} introduced this notion by proving there is 
no non-contextual hidden-variables theory which can reproduce the outcomes predicted by quantum physics.
Here  {\em contextual} means that  the outcome of a measurement on a quantum system depends on the context, i.e. a set of 
compatible measurements (set of mutually commuting observables\footnote{In quantum physics, the outcomes of a measurement are encoded in an hermitian operator, called an observable.
The eigenvalues of the observable correspond to the possible outcomes of the measurement and the  eigenvectors 
correspond to the possible projections of the state after measurement.}) that are performed in the same experiment.
The original proof of Kochen and Specker is based on the impossibility to 
assign coloring (i.e. predefine values 
for the outcomes) to some vector basis associated to some set of projection operators.
Let us present here a simple and nice observable-based proof of the Kochen-Specker Theorem due to D. Mermin \cite{mermin} and A. Peres \cite{peres}. 
Let us denote by $X,Y$ and $Z$ the usual Pauli matrices,
\begin{equation}
 X=\begin{pmatrix}
    0 & 1\\
    1 & 0
   \end{pmatrix},\ Y=\begin{pmatrix}
   0 & -i\\
   i & 0\\
   \end{pmatrix},\ Z=\begin{pmatrix}
   1 & 0\\
   0 & -1
\end{pmatrix}.
\end{equation}

\begin{figure}[!h]
 \begin{center}
    \includegraphics[width=7cm]{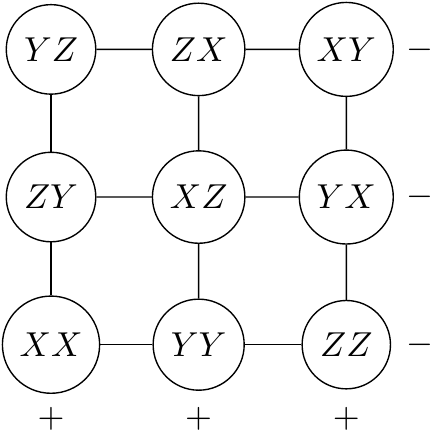}
   \caption{The Mermin-Peres <<Magic>> square.}\label{square}
   \end{center}
\end{figure}

Those three hermitian operators encode the possible measurement outcomes of a spin-$\frac{1}{2}$-particle in a Stern-Gerlach apparatus oriented in three different space directions.
Taking tensor products of two such Pauli matrices we can define Pauli operators acting on two qubits. In \cite{peres,mermin} Mermin and Peres considered 
a set of 2-qubit Pauli
operators similar to the one reproduced in Figure \ref{square}.

This diagram, called the <<Magic>> Mermin-Peres square, furnishes a proof of the impossibility to predict the outcomes of quantum 
physics with a non-contextual hidden-variables theory as I now explain. Each node of the square
represents a 2-qubit observable which squares to identity, i.e. the possible eigenvalues of each node (the possible measurement outcomes) are $\pm 1$. The operators which belong to a row 
or a column are mutually commuting, i.e. they represent a context or a set of compatible observables. The products of each row or column give either $I_4$ or $-I_4$ as indicated by the signs on the diagram.
The odd number of negative rows makes it impossible to pre-assign to each node outcomes ($\pm 1$) which are simultaneously compatible with the constrains on 
the rows (the products of the eigenvalues should be negative)
and columns (the product of the eigenvalues are positive). Therefore, any hidden-variables theory capable of reproducing the outcomes of the measurement that can be achieved 
with the Mermin-Peres
square, should be contextual, i.e. the deterministic values that we wish to assign should be context dependent. This other 
paradox has been studied intensively in the last decade and experiments \cite{amselem,cabello1,bartosik,Kirchmair} 
are now conducted to produce contextuality in the laboratory, leading to consider contextuality as another quantum 
resource for quantum computation or quantum processing \cite{amselem,HWVE}.

Both entanglement of multipartite pure quantum systems and contextual configurations of multi-qubit Pauli observables 
can be nicely described by geometric constructions.
 To put into perspective those two problems and their corresponding 
geometric descriptions, I choose to emphasize  their relations with representation theory. In Section \ref{geometry}, I 
introduce the geometric language of auxiliary varieties and I explain how various classification results
introduced in the quantum information literature in the past 15 years can be uniformly described in terms 
of representation theory. In Section \ref{context}, I describe geometrically 
the set of commutation relations within the $N$-qubit Pauli group and explain 
through explicit examples how weight diagrams of some simple Lie algebras can be extracted from such commutation relations.

\section{The Geometry of Entanglement}\label{geometry} 
In the first part of the paper, I discuss the question of the classification of entanglement  for multipartite quantum 
systems under the SLOCC 
 group from the point of view of algebraic geometry and representation theory. 
In the past fifteen years, 
there have been a lot of papers on the subject tackling the classification for different types of quantum systems 
\cite{Dur,BDD, BDDER,brody2,CDG,CD,DF,Levay1,LV,My,My2,Sarosi}. 
The most famous one  is probably the paper of D\"ur, Vidal and Cirac \cite{Dur} where it was shown that $3$-qubit quantum states 
can be genuinely entangled in two different ways.
\subsection{Entanglement under SLOCC, tensor rank and algebraic geometry}
The Hilbert space of an $n$-partite system will be the tensor product of $n$-vector spaces, where each vector space is the Hilbert space of each individual part. 
Thus the Hilbert space of an $n$-qudit system is  $\mathcal{H}=\CC^{d_1}\otimes\dots\otimes \CC^{d_n}$.
A quantum state being defined up to a phase, we will work in  the projective Hilbert space and denote by $[\psi] \in \PP(\mathcal{H})$ the class of a quantum state $\ket{\psi}\in \mathcal{H}$.
The group of local reversible operations, $G=\SL_{d_1}(\CC)\times\dots\times \SL_{d_n}(\CC)$ acts on $\PP(\mathcal{H})$ 
by its natural action. This group is known in  physics  as the 
group of Stochastic Local Operations with Classical Communications \cite{bennett,Dur} and will be denoted by SLOCC. 

According to the axioms of quantum physics, it would be more
natural to look at entanglement 
classes of multipartite quantum systems under the group of Local Unitary transformations, 
LU$=\SU(d_1)\times\dots\times \SU(d_n)$. In quantum information theory one also considers a larger set of transformations called 
LOCC transformations 
(Local Operations with Classical Communications) which include local unitary and measurement operations (coordinated by classical communications).
Under LOCC two quantum states are equivalent if they can be exactly interconverted by LU operations\footnote{Physically one may imagine that 
each part of the system is in a different location and experimentalists only apply local quantum transformations, i.e. some unitaries defined by local Hamiltonians.}. 
However, the SLOCC equivalence also has a physical meaning as explained in \cite{bennett,Dur}. It corresponds to an equivalence between states that can be interconverted into each other but not with certainty.
Another feature of SLOCC is that if we consider measure of entanglement, the amount of entanglement may increase or decrease under SLOCC while it is invariant under LU and non-increasing under LOCC. However, 
entanglement cannot be created or destroyed by SLOCC and a communication protocol based on a quantum state $\ket{\psi_1}$ can also be achieved with a SLOCC equivalent state $\ket{\psi_2}$ (eventually with different probability of success).
In this sense, SLOCC equivalence is more a qualitative way of separating non equivalent quantum states.

The set of separable, or non-entangled states, is the set of quantum states $\ket{\psi}$ which can be factorized, i.e.
\begin{equation}
 \ket{\psi}=\ket{\psi_1}\otimes\dots\otimes \ket{\psi_n} \text{ with } \ket{\psi_k}\in \CC^{d_k}.
\end{equation}

In algebraic geometry the projectivization of this set is a well-known algebraic variety\footnote{In this paper an algebraic variety will always be
the zero locus of a collection of homogeneous polynomials.} of $\PP(\mathcal{H})$, known as the Segre embedding of 
the product of projective spaces $\PP^{d_1-1}\times\dots\times \PP^{d_n-1}$.

More precisely, let us consider the following map,
\begin{equation}
 \begin{array}{cccc}
    Seg:&  \PP^{d_1-1}\times\dots\times \PP^{d_n-1} & \to & \PP^{d_1\times\dots \times d_n-1}=\PP(\mathcal{H})\\
      & ([\psi_1],\dots,[\psi_n]) & \mapsto & [\psi_1\otimes \dots \otimes \psi_n].
     \end{array}
\end{equation}

The image of this map is the Segre embedding of the product of projective spaces and clearly coincides with $X_\Sep$, the projectivization of the set of separable states.  We will thus write
\begin{equation} X_{\Sep}=\PP^{d_1-1}\times \dots\times \PP^{d_n-1}\subset \PP(\mathcal{H}).\end{equation}
 The Segre variety has the property to be the only one closed orbit of $\PP(\mathcal{H})$ for the  $\SLOCC$ action.
Up to local reversible transformations, every separable state $\ket{\psi}=\ket{\psi_1}\otimes \dots\otimes \ket{\psi_n}$ can 
be transformed to 
$\ket{0}\otimes \dots\otimes\ket{0}=\ket{0\dots 0}$ if we assume that each vector space $\CC^{d_i}$ is equipped with a 
 basis denoted by $\ket{0},\dots,\ket{d_i-1}$,
\begin{equation}
 X_{\Sep}=\PP^{d_1-1}\times \dots \times\PP^{d_n-1}=\PP(\SLOCC.\ket{0\dots 0})\subset \PP(\mathcal{H}).
\end{equation}

A quantum state $\ket{\psi}\in \mathcal{H}$ is entangled iff it is not separable, i.e. 
\begin{equation}\ket{\psi} \text{ entangled }\Leftrightarrow [\psi]\in \PP(\mathcal{H}\setminus X_{\Sep}).\end{equation}

In algebraic geometry, it is usual to study properties of $X$ by introducing auxiliary varieties, i.e. 
varieties built from the knowledge of $X$, whose attributes (dimension, degree) will 
tell us something about the geometry of $X$.

Let us first introduce two auxiliary varieties of importance for quantum information and entanglement: the secant and tangential varieties.

\begin{definition}
 Let $X\subset \PP(V)$ be a projective algebraic variety, the secant variety of $X$ is the Zariski closure of the union of secant lines, i.e.
 \begin{equation}
  \sigma_2(X)=\overline{\cup_{x,y\in X}\PP^1 _{xy}},
 \end{equation}
 where $\PP^1_{xy}$ is the projective line corresponding to the projectivization of the linear span $\text{Span}(\hat{x},\hat{y})\subset V$ (a $2$-dimensional linear subspace of $V$).
\end{definition}
\begin{remark}\rm
 This definition can be extended to higher-dimensional secant varieties. More generally, one may define the $k$th-secant variety of $X$, 
 \begin{equation}
  \sigma_k(X)=\overline{\cup_{x_1,\dots,x_k} \PP_{x_1,\dots,x_k}^{k-1}},
 \end{equation}
 where now $\PP^{k-1}_{x_1,\dots,x_k}$ is the a projective subspace of dimension $k-1$ obtained as the projectivization of the linear span $\text{Span}(\hat{x}_1,\dots,\hat{x}_n)\subset V$. 
 There is a natural sequence of inclusions given by $X\subset \sigma_2(X)\subset \sigma_3(X)\subset \dots\subset \sigma_q(X)=\PP(V)$, where $q$ is the smallest integer such that the $q$th-secant variety fills the ambient space.
\end{remark}

\begin{remark}\rm
 The notion of secant varieties is deeply connected to the notion of rank of tensors. One says that a tensor $T\in \CC^{d_1}\otimes \dots\otimes  \CC^{d_n}$
 has rank $r$ iff $r$ is the smallest integer such that $T=T_1+\dots+ T_r$ and each tensor $T_i$ can be factorized, i.e. $T_i=a^i_{1}\otimes\dots\otimes a^i_n$. From the definition one sees 
 that the Segre variety $\PP^{d_1-1}\times\dots \times \PP^{d_n-1}$ corresponds to the projectivization of rank-one tensors of $\mathcal{H}$ and the secant variety 
 of the Segre is the Zariski closure of the (projectivization of) rank-two tensors because a generic point of $\sigma_2(\PP^{d_1-1}\times\dots \times \PP^{d_n-1})$  is the sum of two rank-one tensors.
 Similarly, $\sigma_k(\PP^{d_1-1}\times\dots \times \PP^{d_n-1})$ is the algebraic closure of the set of rank at most $k$ tensors. Tensors (states) which belong to 
 $\sigma_k(\PP^{d_1-1}\times\dots \times \PP^{d_n-1})\backslash \sigma_{k-1}(\PP^{d_1-1}\times\dots \times \PP^{d_n-1})$ will be called tensors (states) of border rank-$k$, i.e. they can 
 be expressed as (limits) of rank $k$ tensors.
\end{remark}

Another auxiliary variety of importance is the tangential variety, i.e. the union of tangent spaces. When $x\in X$ is a smooth point 
of the variety I denote by $T_x X$ the projective tangent space and $\hat{T}_x X$ its cone in $\mathcal{H}$.
\begin{definition}
 Let $X\subset \PP(V)$ be a smooth projective algebraic variety, the tangential variety of $X$ is defined by
 \begin{equation}
  \tau(X)=\cup_{x\in X} T_x X,
 \end{equation}
 (here the smoothness of $X$ implies that the union is closed).
\end{definition}

The auxiliary varieties built from $X_{\Sep}$ are of importance to understand the entanglement stratification of Hilbert spaces of pure quantum systems under SLOCC for mainly two reasons. 
First the auxiliary varieties 
are SLOCC invariants by construction because $X_{\Sep}$ is a SLOCC-orbit. Thus the construction of auxiliary
varieties from the core set of separable states $X_{\Sep}$ produces a stratification of 
the ambient space by SLOCC-invariant algebraic varieties. The possibility to stratify the ambient 
space by  secant varieties was known to geometers more than a century ago \cite{Terra}, but it was noticed to be useful for studying entanglement classes only 
recently by Heydari \cite{Hey}.
It is equivalent to a stratification of the ambient space by the (border) ranks of the states which, as pointed out by 
Brylinski, can be considered as an algebraic measure of 
entanglement \cite{Bry}. 

The second interesting aspect of those auxiliary varieties, in particular the secant and tangent one, is that they may have a nice quantum information interpretation. To  be more precise, let us recall the definition of 
the $\ket{GHZ_n}$ and $\ket{W_n}$ states,
\begin{equation}
 \ket{GHZ_n}=\dfrac{1}{\sqrt{2}}(\ket{0\dots 0}+\ket{1\dots 1}),
\end{equation}

\begin{equation}
 \ket{W_n}=\dfrac{1}{\sqrt{n}}(\ket{100\dots 0}+\ket{010\dots 0}+\dots+\ket{00\dots 1}).
\end{equation}

Then we have the following geometric interpretations of the closure of their corresponding SLOCC classes,

\begin{equation}
\overline{\SLOCC.[GHZ_n]}=\sigma_2(X_{\Sep}) \text{ and } \overline{\SLOCC.[W_n]}=\tau(X_{\Sep}).
\end{equation}

It is not difficult to see why the Zariski closure of the SLOCC orbit of the $\ket{GHZ_n}$ state is 
the secant variety of the set of separable states. Recall that a generic point
of $\sigma(X_{\Sep})$ is a rank $2$ tensor. Thus, if $[z]$ is a generic point of $\sigma_2(X_{\Sep})$, one has
\begin{equation}
 [z]=[\lambda x_1\otimes x_2\otimes \dots\otimes x_n+\mu y_1\otimes y_2 \otimes\dots \otimes y_n],
\end{equation}
with $x_i, y_i \in \CC^{d_i}$. Because $[z]$ is generic we may assume that $(x_i,y_i)$ are linearly independent. Therefore, 
there exists $g_i\in \SL_{d_i}(\CC)$ such that $g_i.x_i\propto\ket{0}$ and 
$g_i.y_i\propto\ket{1}$ for all $i\in\{1,\dots,n\}$. Thus we can always find $g\in \SLOCC$ such that $[g.z]=[GHZ_n]$.

To see why the tangential variety of the variety of separable states always corresponds to the (projective) orbit closure of the $\ket{W_n}$ state, we need to show that a generic tangent 
vector of $X_{\Sep}$ 
is always SLOCC equivalent to $\ket{W_n}$. A tangent vector can be obtained by differentiating a curve of $X_\Sep$. Let 
$[x(t)]=[x_1(t)\otimes x_2(t)\otimes\dots\otimes x_n(t)]\subset X_{\Sep}$ with 
$[x(0)]=[x_1\otimes x_2\otimes\dots\otimes x_n]$. Because we are looking at a generic tangent vector, we assume 
that for all $i$, $x_i'(0)=u_i$ and $u_i$ is not collinear to $x_i$. 
Then Leibniz's rule insures that
\begin{equation}
 [x'(0)]=[u_1\otimes x_2\otimes\dots \otimes x_n+x_1\otimes u_2\otimes \dots \otimes x_n+\dots+x_1\otimes x_2\otimes \dots \otimes u_n].
\end{equation}

Let us consider $g_i\in \SL_{d_i}(\CC)$ such that $g_i.x_i\propto\ket{0}$ and $g_i.u_i\varpropto\ket{1}$, then we obtain 
$[g.x'(0)]=[W_n]$ for $g=(g_1,\dots,g_n)$.

An important result regarding the relationship between tangent and secant varieties is due to Fyodor Zak \cite{Zak}.
\begin{theorem}[\cite{Zak}]\label{Zak}
Let $X\subset \PP(V)$ be a projective algebraic variety of dimension $d$. Then one of the following two properties holds,
\begin{enumerate}
 \item $dim(\sigma_2(X))=2d+1$ and $dim(\tau(X))=2d$,
 \item $dim(\sigma_2(X))\leq 2d$ and $\tau(X)=\sigma_2(X)$.
\end{enumerate}
\end{theorem}

To get information  from Zak's theorem one needs to compute the dimension of the secant variety of $X$. This can be done by using
 an old geometrical result from the beginning of the XXth century known as Terracini's Lemma.

\begin{lemma}[Terracini's Lemma]
  Let $[z]\in \sigma_2(X)$ with $[z]=[x+y]$ and $([x],[y])\in X\times X$ be a general pair of points. Then
 \begin{equation}
  \hat{T}_{[z]} \sigma_2(X)=\hat{T}_{[x]}X+\hat{T}_{[y]} X.
 \end{equation}
\end{lemma}

Terracini's Lemma tells us that if $X$ is of dimension $d$, 
 the expected dimension of $\sigma_2(X)$ is $2(d+1)-1=2d+1$. 
Thus by Zak's Theorem, one knows that if $\sigma_2(X)$ has the expected dimension then the tangential variety is a proper subvariety of $\sigma_2(X)$ and otherwise both varieties are the same.

\begin{ex}\label{secant3qubit}
Let us look at the case where $X_\Sep=\PP^1\times\PP^1\times \PP^1\subset\PP^7$.  The dimension of $\sigma_2(X_\Sep)$ can be obtained as a simple application of Terracini's lemma. 
Let $[x]=[\phi_1\otimes\phi_2\otimes \phi_3]\in X_{\Sep}$ then $\hat{T}_{[x]} X_\Sep=\CC^2\otimes \phi_2\otimes \phi_3+\phi_1\otimes  \CC^2\otimes \phi_3+\phi_1\otimes \phi_2\otimes  \CC^2$.
Thus one gets for $[GHZ]=[\ket{000}+\ket{111}]\in \sigma_2(X_\Sep)$,
\begin{equation}
\begin{array}{lll}
 \hat{T}_{[GHZ]} \sigma_2(X_\Sep)& = &\hat{T}_{[\ket{000}]} X_\Sep+\hat{T}_{[\ket{111}]}  X_{\Sep}\\
 &= &\CC^2\otimes \ket{0}\otimes\ket{0}+\ket{0}\otimes \CC^2\otimes\ket{0}+ \ket{0}\otimes\ket{0}\otimes \CC^2\\
 & &+\CC^2\otimes \ket{1}\otimes\ket{1}+\ket{1}\otimes \CC^2\otimes\ket{1}+ \ket{1}\otimes\ket{1}\otimes \CC^2.
\end{array}
 \end{equation}
 Therefore $\dim(\hat{T}_{[GHZ]} \sigma_2(X_\Sep))=8$, i.e. $\dim(\sigma_2(X_\Sep))=7$.
\end{ex}

\subsection{The three-qubit classification via auxiliary varieties}\label{tripartitesec}
As mentioned at the beginning of the section, the problem of the classification of multipartite 
quantum systems acquired a lot of attention after 
D\"ur, Vidal and Cirac's paper \cite{Dur} on the classification of three-qubit states, where it was first shown that two quantum states 
can be entangled in two genuine non-equivalent ways. The authors showed that for three-qubit systems there are exactly 
$6$ SLOCC orbits whose representatives can be chosen to be: $\ket{Sep}=\ket{000}$, $\ket{B_1}=\dfrac{1}{\sqrt{2}}(\ket{000}+\ket{011})$, 
$\ket{B_2}=\dfrac{1}{\sqrt{2}}(\ket{000}+\ket{101})$, $\ket{B_3}=\dfrac{1}{\sqrt{2}}(\ket{000}+\ket{110})$, $\ket{W_3}=\dfrac{1}{\sqrt{3}}(\ket{100}+\ket{010}+\ket{001})$ and $\ket{GHZ_3}=\dfrac{1}{\sqrt{2}}(\ket{000}+\ket{111})$.

The state $\ket{Sep}$ is a representative of the orbit of separable states and the states $\ket{B_i}$ are bi-separable. The only 
genuinely entangled states are $\ket{W_3}$ and $\ket{GHZ_3}$. It turns out that this orbit classification of the  Hilbert space 
of three qubits was known long before the famous paper of D\"ur, Vidal and Cirac from  different mathematical 
perspectives (see for example \cite{Pav,GKZ}). Probably the oldest mathematical proof of this result goes back to the work of Le Paige (1881) who classified 
the trilinear binary forms under (local) linear transformations in \cite{LePai}.

From a geometrical point of view the existence of two distinguished orbits corresponding to $\ket{W_3}$ and $\ket{GHZ_3}$ can be obtained 
as a consequence of Zak's theorem (Theorem \ref{Zak}).  Indeed, in Example \ref{secant3qubit} one shows that the secant variety 
of the variety of separable three qubit states has the expected dimension and fills the ambient space.
 According to Zak's Theorem, this implies that 
 the tangential variety $\tau(X_\Sep)$ is a codimension-one sub-variety of $\sigma_2(X_\Sep)=\PP^7$ and, therefore, both 
 orbits are distinguished.
In other words, from a geometrical perspective there exist two non-equivalent, genuinely entangled states for the 
three-qubit system because the secant variety of the set of separable states
has the expected dimension and fills the ambient space.

\begin{figure}[!h]
\begin{center}
 \includegraphics{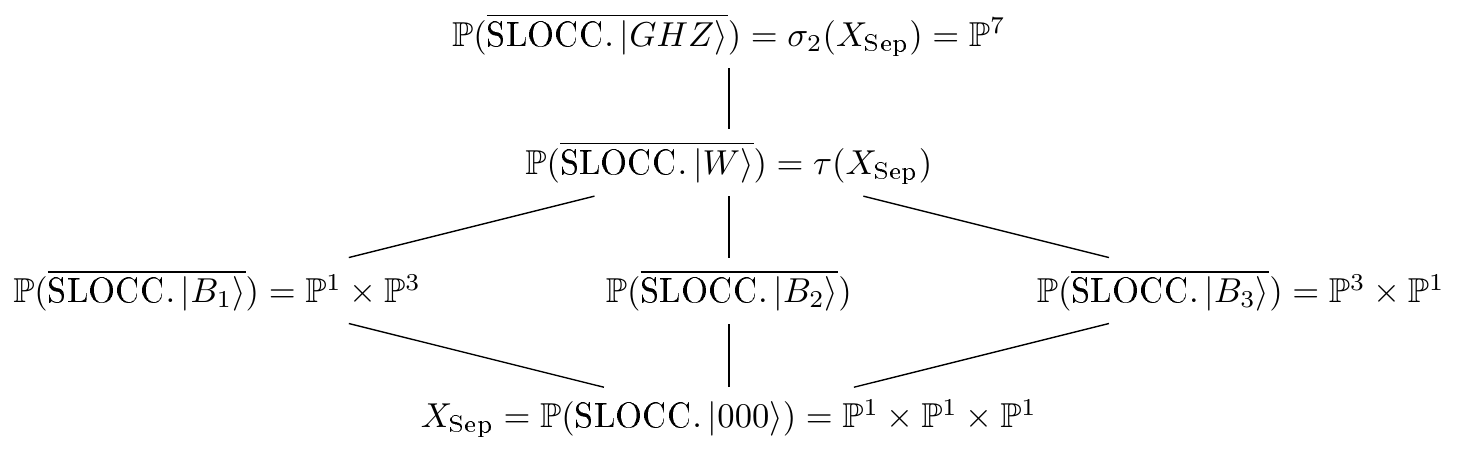}
 \caption{Stratification of the (projectivized) Hilbert space of three qubits by SLOCC-invariant algebraic varieties (the secant and tangent).}\label{222}
\end{center}
 \end{figure}

In this language of auxiliary varieties let us also mention that the orbit closures defined by the bi-separable states $\ket{B_i}$
have also a geometric interpretation. For instance, $\ket{B_1}=\ket{0}\otimes \frac{1}{\sqrt{2}}(\ket{00}+\ket{11})=\ket{0}\otimes \ket{EPR}$.
The projective orbit closure is 
\begin{equation}
 \PP(\overline{\SLOCC.\ket{B_1}})=\PP^1\times\PP^3\subset \PP^7,
\end{equation}
where $\PP^3=\sigma_2(\PP^1\times  \PP^1)$.
The geometric stratification  by SLOCC invariant algebraic varieties in the $3$-qubit case can be represented as in Fig. \ref{222}.


\begin{remark}\rm
 This idea of introducing auxiliary varieties  to describe SLOCC classes of entanglement also appears in \cite{sanz,Sawicki}. It allows one to connect the study 
 of entanglement in quantum information to a large literature in mathematics, geometry and their applications. 
For instance, the question of finding defining equations 
 of auxiliary varieties is central in many areas of applications of mathematics 
 to computer science, signal processing or phylogenetics (see the introduction  of \cite{Landsberg} and  references therein).
 Those equations can be obtained by mixing techniques from representation theory and geometry \cite{LM1,LO,O1}.
 In the context of quantum information finding defining equations of auxiliary varieties provides tests 
 to decide if two states could be SLOCC equivalent.
 Classical invariant theory also provides tools to generate invariant and covariant polynomials 
 \cite{BLT1,BLT2,LT1,LT2} and these techniques were used in \cite{HLT,HLT2,HLT3} to identify  entanglement classes with auxiliary varieties.
\end{remark}

\subsection{Geometry of hyperplanes: the dual variety}
Another auxiliary variety of interest is the dual variety of $X_\Sep$:
 \begin{equation}
  X_\Sep^*=\overline{\{ H\in (\PP^N)^*, \exists x\in X_\Sep, T_x X_\Sep\subset H\}}.
 \end{equation}
 The variety $X_\Sep^*$ parametrizes the set of hyperplanes defining  singular (non-smooth) hyperplane sections of $X_\Sep$. Using the hermitian inner product on $\mathcal{H}$, one can identify the dual variety
 of $X_\Sep$ with the set of states which define a singular hyperplane section  of $X_\Sep$. More precisely, given
 a state $\ket{\psi}\in \mathcal{H}$ we have 
 \begin{equation}\ket{\psi}\in X_\Sep^* \text{ iff }X_\Sep\cap H_\psi=\{\ket{\varphi}\in X_\Sep, \langle \psi,\varphi\rangle=0\} \text{ is singular}.\end{equation}
 For $X_\Sep=\PP^{d_1-1}\times \PP^{d_2-1}\times \dots \times \PP^{d_n-1}$ (with $d_j\leq \sum_{i\neq j} d_i$),  the 
variety $X_\Sep^*$ is always a hypersurface, called the hyperdeterminant of format $d_1\times d_2\times\dots\times d_n$ 
\cite{GKZ}.  By construction the hyperdeterminant is SLOCC-invariant and so is its  singular locus. Therefore, the hyperdeterminant and its singular locus 
can be used to stratify the (projectivized) Hilbert space under SLOCC.

\begin{figure}[!h]
\begin{center}
 \includegraphics{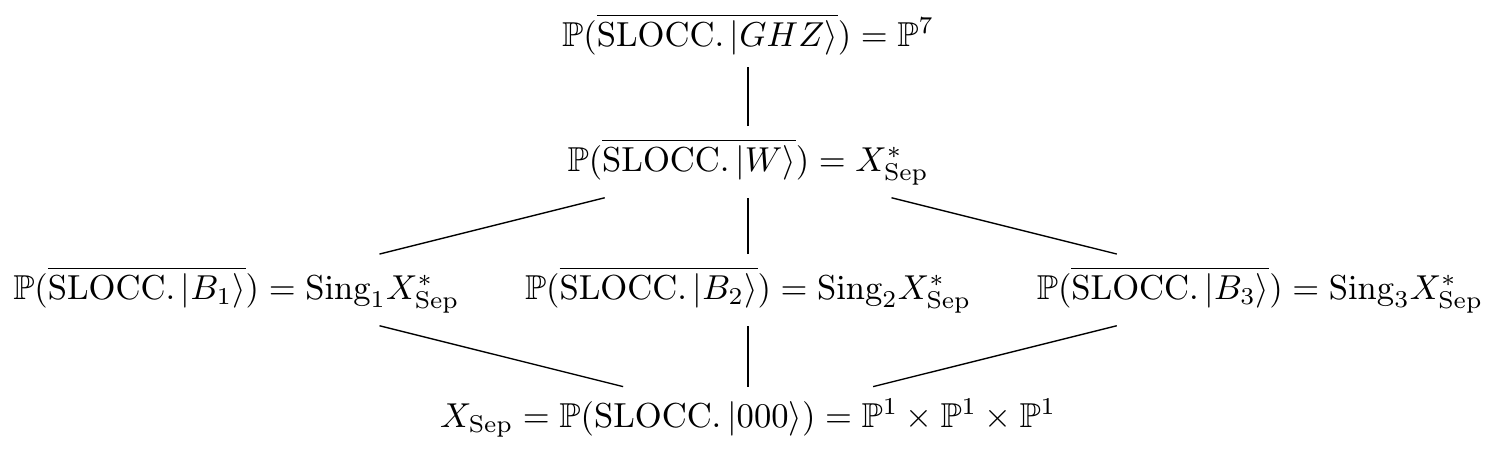}
 \caption{Stratification of the (projectivized) Hilbert space of three qubit by SLOCC-invariant algebraic varieties (the dual and its singular locus). $\text{Sing}_i X_\Sep ^*$ represent different components of the singular locus \cite{My,WZ}.}\label{3qubitdual}
\end{center}
 \end{figure}

This idea goes back to Miyake \cite{My,My2,My3} who interpreted  previous 
results of Weyman and Zelevinksy on singularities of hyperdeterminants \cite{WZ} to describe the entanglement structure 
for the $3$- and $4$-qubit systems, as well as for the $2\times 2\times n$-systems. Following Miyake, the hyperdeterminant of format 
$2\times 2\times 2$, also known as the Cayley hyperdeterminant, provides a dual picture of the 
three-qubit classification (Fig. \ref{3qubitdual}).

One can go further by studying which types of singular  hyperplane sections can be associated to a given state.

To do so we use the rational map defining the Segre embedding to obtain the equations of the hyperplane sections:

\begin{equation}\label{segre2}
 \begin{array}{ccc}
   \PP^{d_1-1}\times\dots \times \PP^{d_n-1} & \to & \PP(\mathcal{H}) \\
       ([x^1 _1:\dots:x^1 _{d_1}],\dots,[x_1 ^n:\dots:x_{d_n} ^n]) & \mapsto & [x^1 _1 x^2_1\dots x^n _1:\dots: {\bf x}_J:\dots:x^1 _{d_1}x^2_{d_2}\dots x_{d_n} ^n ],
 \end{array}
 \end{equation}
 where ${\bf x}_J$, for $J=(i_1,\dots,i_n)$ with $1\leq i_j\leq d_j$, denotes the monomial ${\bf x}_J=x^1_{i_1}x^2 _{i_2}\dots x^n _{i_n}$. In (\ref{segre2}) the monomials ${\bf x}_J$ 
 are ordered 
 lexicographically in terms of  multi-indices $J$.
Therefore to a state $\ket{\psi}=\sum a_{i_1\dots i_n}\ket{i_1\dots i_n}$ one associates the hypersurface of $X_{\Sep}$ defined by 
\begin{equation}
 f_{\ket{\psi}}=\sum_{i_1,\dots,i_n} a_{i_1\dots i_n} x^1 _{i_1} \dots x^n _{i_n}=0.
\end{equation}
If $\ket{\psi} \in X_\Sep^*$, then $f_{\ket{\psi}}$ is a singular homogeneous polynomial, i.e.
 there exists $\overline{x}\in X_\Sep$ such that
 \begin{equation}
  f_{\ket{\psi}}(\overline{x})=0 \text{ and } \partial_{i_k} f_{\ket{\psi}}(\overline{x})=0.
 \end{equation}
 In the 70s Arnol'd defined and classified {\em simple singularities} of complex functions \cite{Ar,Ar2}. 
 \begin{definition}
 One says that $(f_{\ket{\psi}},\overline{x})$ is simple iff under a small perturbation it can only degenerate to a 
 finite number of non-equivalent singular hypersurfaces $(f_{\ket{\psi}}+\varepsilon g,\overline{x}')$ (up to biholomorphic change of coordinates).
 \end{definition}
 Simple singularities are always isolated, i.e. the 
 Milnor number $\mu=\text{dim}\CC[x_1,\dots,x_n]/(\nabla f_{\overline{x}})$ is finite, and they can be classified in 5 families (Table \ref{simple}).

 \begin{table}[!h]
 \begin{center}
 \begin{tabular}{c|c|c|c|c|c}
 {Type} & $A_k$ & $D_k$ & $E_6$ & $E_7$ & $E_8$ \\
  \hline
 Normal form & $x^{k+1}+y^2$& $x^{k-1}+ xy^2$& $x^3+y^4$ & $x^3+xy^3$ & $x^3+y^5$\\
  \hline
 {Milnor number}& $k$ & $k$ & $6$ & $7$ & $8$
 \end{tabular}
 \caption{Simple singularities and their normal forms.}\label{simple}
\end{center}
 \end{table}
 
\begin{figure}[!h]
 \begin{center}
 \includegraphics[width=6cm]{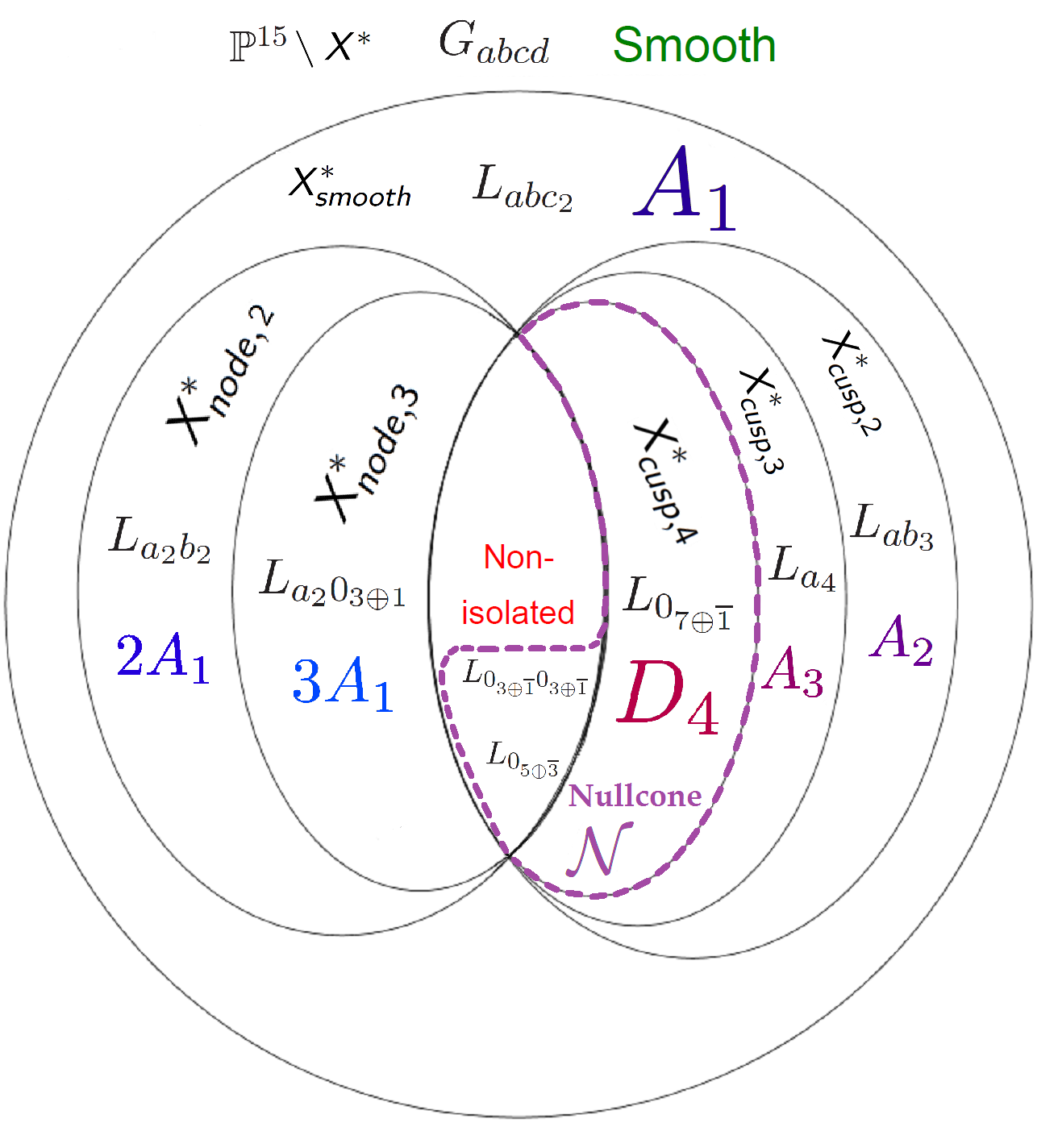}
  \includegraphics[width=6.3cm]{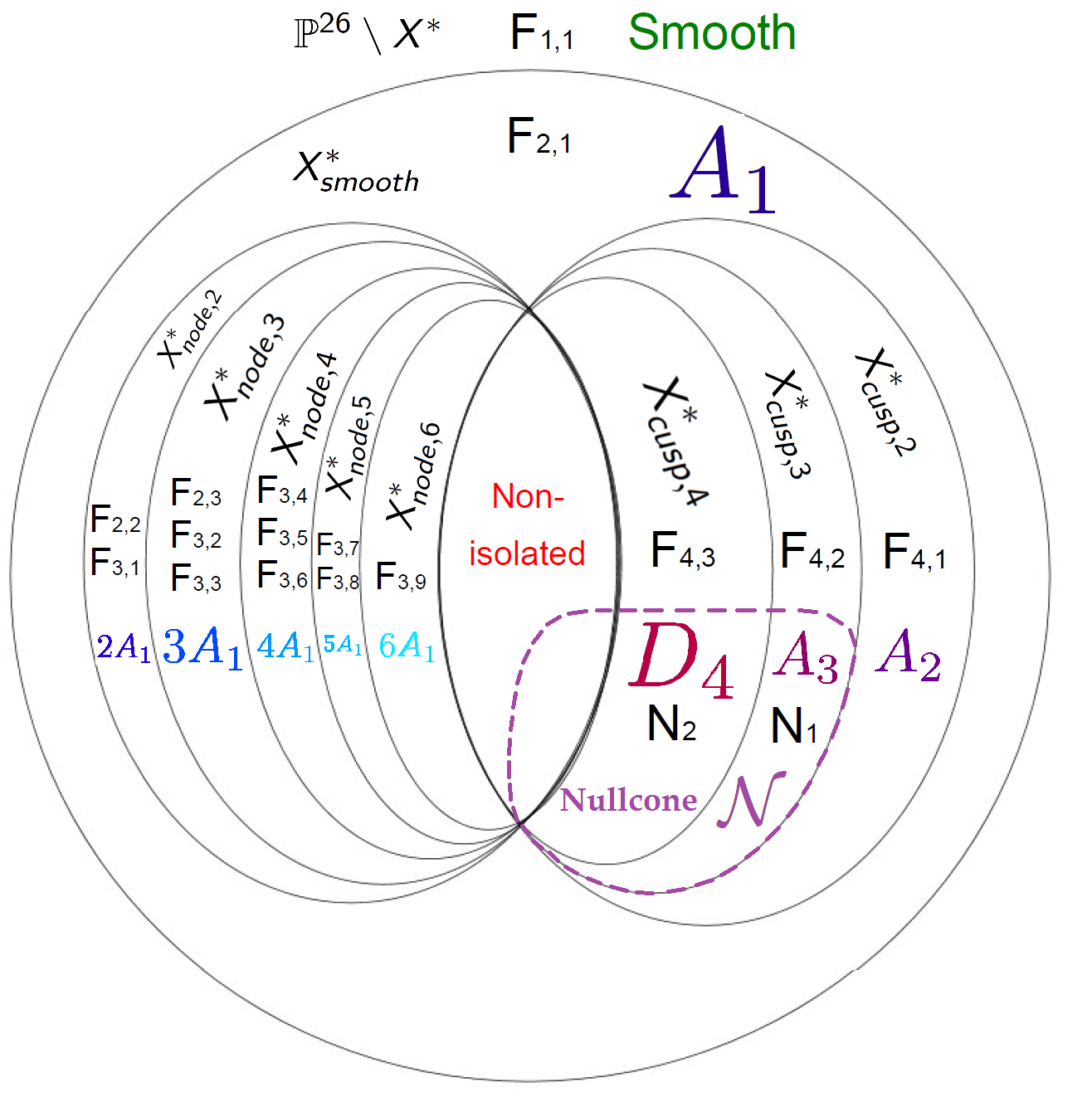}
  \caption{Four-qubit and three-qutrit entanglement stratification by singular types of the hyperplane sections. Thus cusp components correspond to states with sigularities 
  which are not of type $A_1$ and the node components correspond to states with at least two singular points \cite{WZ}. The names of the normal forms come from \cite{V} and \cite{Nu}.}\label{onion}
 \end{center}
 \end{figure}

The singular type can be identified by computing the Milnor number, the corank of the Hessian and the cubic 
term in the degenerate directions.
\begin{ex}
Let us consider the 4-qubit state $|\psi\rangle=|0000\rangle+|1011\rangle+|1101\rangle+|1110\rangle$. The parametrization of the variety 
 of separable states is given by $\phi([x_0:x_1],[y_0:y_1],[z_0:z_1],[t_0:t_1])=[x_0y_0z_0t_0:\dots:x_1y_1z_1t_1]$. 
 The homogeneous polynomial associated to $\ket{\psi}$ is 
 \begin{equation} f_{\ket{\psi}}=x_0y_0z_0t_0+x_1y_0z_1t_1+x_1y_1z_0t_1+x_1y_1z_1t_0.\end{equation} 
In the chart $x_0=y_1=z_1=t_1=1$ one obtains locally the hypersurface defined by \begin{equation} f(x,y,z,t)=yzt+xy+xz+xt.\end{equation}
 
The point $(0,0,0,0)$ is the only singular point of  $f_{\ket{\psi}}$ (the hyperplane section is tangent to $[|0111\rangle]$).
 The  Hessian matrix of this singularity has co-rank $2$ and  $\mu=4$.  Therefore the hyperplane section defined by 
 $\ket{\psi}$ has a unique singular point of type $D_4$ and this is true for all states $\SLOCC$ equivalent to $\ket{\psi}$.   
 \end{ex}
 The four-qubit and three-qutrit pure quantum systems are examples of systems with an infinite number of SLOCC-orbits. However, in both
 cases the orbit structure can still be described 
 in terms of family of normal forms by introducing parameters. The $4$-qubit classification was originally obtained 
 by Verstraete et al. \cite{V}  with a small correction provided by \cite{CD}. Regarding the 3-qutrit classification, it has 
 not been published in the quantum physics 
 literature, but it can be directly translated from the orbit classification of the $3\times 3\times 3$ complex hypermatrices under $\text{GL}_3(\CC)\times \text{GL}_3(\CC)\times \text{GL}_3(\CC)$ obtained 
 by Nurmiev \cite{Nu}.  In \cite{HLP, HJ} I calculated with my co-authors the type of isolated singularities associated to those forms.
 First of all, all isolated singularities are simple but moreover the worst, in terms of degeneracy, 
 isolated singularity that arises is, in both cases, of type $D_4$. This allows us to get a more precise onion-like description \cite{My} of the classification, see Fig \ref{onion}. 
  It also gives information  about how a state can be perturbed to another one. For instance,
 for a sufficiently small perturbation a state corresponding to a singular hyperplane section with only isolated singularities can only be changed to a state with isolated singularities 
 of a lower degeneracy.
 \begin{theorem}[\cite{HLP}]\label{mainthm}
 Let $H_\psi$ be a hyperplane of $\PP(\mathcal{H})$ tangent to $X_\Sep=\PP^1\times\PP^1\times\PP^1\times\PP^1\subset \PP^{15}$ and such that $X_\Sep\cap H_\psi$ has 
 only isolated singular points. Then 
 the singularities are either of types  $A_1, A_2, A_3$, or of type $D_4$, and there exist hyperplanes realizing each type of singularity.
 Moreover, if we denote by $\widehat{X}_\Sep^*\subset \mathcal{H}$ the cone over the dual variety of $X_\Sep$, {i.e.} the zero locus of the Cayley
 hyperdeterminant of format $2\times 2\times 2\times 2$, then the quotient map\footnote{In the four-qubit case, the ring of SLOCC invariant polynomials is generated by four polynomials denoted by $H,L,M$ and $D$ in \cite{LT1}. One way of 
 defining the quotient map is to consider $\Phi:\mathcal{H}\to \CC^4$ defined by $\Phi(\hat{x})=(H(\hat{x}),L(\hat{x}),N(\hat{x}),D(\hat{x}))$, see \cite{HLP}.} $\Phi:\mathcal{H}\to \CC^4$ is such that $\Phi(\widehat{X}_\Sep^*)=\Sigma_{D_4}$, where
 $\Sigma_{D_4}$ is the discriminant of the miniversal deformation\footnote{The discriminant of the miniversal deformation of a singularity parametrizes all singular deformations 
 of the singularity \cite{Ar}.} of the $D_4$-singularity.
\end{theorem}

\begin{theorem}[\cite{HJ}]\label{proposition1}
		Let $H_\psi \cap X$ be a singular hyperplane section of the algebraic variety of separable states for three-qutrit systems, 
		i.e. $X_\Sep = \PP^2 \times \PP^2 \times \PP^2\subset \PP^{26}$ defined by a quantum pure state $\ket{\psi}\in \PP^{26}$. Then $H_\psi\cap X_\Sep$ only admits simple or nonisolated singularities. Moreover if  $x$ is an 
		isolated singular point of $H_\psi \cap X_\Sep$,  then its singular type is either $A_1$, $A_2$, $A_3$ or $D_4$.
\end{theorem}
 

\subsection{Representation theory and quantum systems}
Let us now consider $G$, a complex semi-simple Lie group, and $V$, an irreducible representation of $G$, i.e. one considers a map $\rho:G\to \text{GL}(V)$ defining an action of $G$
on $V$ such that there is no proper subspace of $V$ stablized by  $G$. The projectivization of an irreducible 
representation $\PP(V)$ always contains a unique closed orbit $X_G\subset \PP(V)$ called the highest weight 
orbit \cite{FH}. 
 The Hilbert space $\mathcal{H}=\CC^{d_1}\otimes\dots \otimes \CC^{d_n}$
is an irreducible representation of $\SLOCC=\SL_{d_1}(\CC)\times \dots\times \SL_{d_n}(\CC)$ and, in this 
particular case, the highest weight orbit is nothing but the Segre variety $X_\Sep=\PP^{d_1-1}\times\dots\times\PP^{d_n-1}$.

It is natural to ask if other semi-simple Lie groups and representations have physical interpretations in terms of quantum systems.
Let us first introduce the case of symmetric and skew-symmetric states.
\begin{itemize}
 \item Consider the simple complex Lie group  $\SLOCC=\SL_n(\CC)$ and its irreducible representation 
 $\mathcal{H}_{bosons}=\text{Sym}^k(\CC^n)$ where $\text{Sym}^k(\CC^n)$ is the $k$th symmetric tensor product of $\CC^n$. Then $\mathcal{H}_{bosons}$ is the Hilbert space 
 of $k$ indistinguishable symmetric particles, each particle being an $n$-single particle state. Physically, it corresponds to $k$ bosonic 
 $n$-qudit states. Geometrically, the highest weight orbit is the so-called 
 Veronese embedding of $\PP^{n-1}$ \cite{Ha}:
 \begin{equation}
  \begin{array}{llll}
    v_k: &  \PP^{n-1} & \to & \PP(\text{Sym}^k(\CC^n))\\
       & [\psi] & \mapsto & [\underbrace{\psi\circ \psi\circ\dots\circ \psi}_{k \text{ times}}].
      \end{array}
 \end{equation}
The variety $v_k(\PP^{n-1})\subset \PP(\text{Sym}^k(\CC^n))$ is geometrically the analogue of the variety of separable states for multiqudit systems
given by the Segre embedding. It is not completely clear 
what entanglement physically means for 
bosonic systems.  The ambiguity comes from the fact that symmetric states like $\ket{W_3}=\dfrac{1}{3}(\ket{100}+\ket{010}+\ket{001})$ can be factorized 
under the symmetric tensor product $\ket{W_3}=\ket{1}\circ \ket{0}\circ \ket{0}$. However 
we can define  entanglement in such symmetric systems
by considering the space of symmetric states as a subset of the space of $k$ $n$-dits states 
$\CC^n\otimes \dots \otimes \CC^n$. In this case the $k$th-Veronese embedding of $\PP^{n-1}$ corresponds to the intersection of the 
variety of separable states $\PP^{n-1}\times \dots \times\PP^{n-1}$ with $\PP(\text{Sym}^k(\CC^n))$ \cite{brody2}.
In the special case of $n=2$, the variety $v_k(\PP^1)\subset \PP^{k}$ can also be identified with the variety of spin $s$-coherent states ($2s=k$)
 when a spin $s$-state is given as a collection of $2s$ spin $\frac{1}{2}$-particles \cite{chrys,Aulbach2}.
For a comprehensive study about entanglement of symmetric states, see \cite{Aulbach}.
 
 \item Consider the simple complex Lie group $\SLOCC=\SL_n(\CC)$ and its irreducible representation 
 $\mathcal{H}_{fermions}=\bigwedge^k \CC^n$ which is the the space of skew symmetric $k$ tensors over $\CC^n$. This Hilbert space represents the space of $k$ skew-symmetric particles with $n$-modes, i.e. $k$ fermions 
 with $n$-single particle states. In this case the highest weight orbit is also a well-known algebraic 
 variety, called the Grassmannian variety $G(k,n)$. The Grassmannian variety $G(k,n)$ is the set of $k$ planes in $\CC^n$ and 
 it is defined as a subvariety of $\PP(\bigwedge^k \CC^n)$ by the Pl\"ucker embedding \cite{Ha}:
 \begin{equation}
  \begin{array}{lll}
   G(k,n) & \hookrightarrow & \PP(\bigwedge^k \CC^n)\\
   \text{Span}\{v_1,v_2,\dots,v_k\} & \mapsto & [v_1\wedge v_2\wedge \dots \wedge v_k].
  \end{array}
 \end{equation}
From the point of view of quantum physics the Grassmannian variety represents the set of fermions with Slater rank one
and is naturally considered as the set of non-entangled states.
\end{itemize}

Another type of quantum system which can be described by means of representation theory is the case 
of particles in a fermionic Fock space with finite $N$-modes \cite{Sarosi}. A fermionic Fock space with finite $N$-modes 
physically describes 
fermionic systems with $N$-single particle states, where the number of particles is not necessarily 
conserved by the admissible transformations. Let us recall the basic ingredient to describe such a Hilbert space.
Let $V$ be an $N=2n$-dimensional complex vector space corresponding to one particle states. The associated fermionic Fock space is given by: 
 
 \begin{equation}
 \mathcal{F}=\wedge^{\bullet} V=\CC\oplus V\oplus \wedge^2 V\oplus \dots \oplus \wedge ^N V=\underbrace{\wedge^{even} V}_{\mathcal{F}_{+}}\oplus \underbrace{\wedge^{odd} V}_{\mathcal{F}_{-}}.
 \end{equation}
 
 Similarly to the bosonic Fock space description of the Harmonic oscillator, one may describe this
 vector space as generated from the vacuum $\ket{0}$ (a generator of $\wedge^0 V$) by applying 
 creation operators ${\bf p}_i$, $1\leq i\leq N$. Thus
 a state $\ket{\psi} \in \mathcal{F}$ is given by 
 \begin{equation}
  \ket{\psi}=\sum_{i_1,\dots,i_k} \psi_{i_1,\dots, i_k} {\bf p}_{i_1}\dots {\bf p}_{i_k} \ket{0} \text{ with } \psi_{i_1,\dots,i_k}  \text{ skew symmetric tensors}.  
 \end{equation}
The annihilation operators ${\bf n}_j$, $1\leq j\leq N$ are defined such that 
${\bf n}_j\ket{0}=0$ and satisfy the Canonical Anticommutation Relations (CAR)
\begin{equation}
 \{{\bf p}_i,{\bf n}_j\}={\bf p}_i{\bf n}_j+{\bf n}_j{\bf p}_i=\delta_{ij}, \{{\bf p}_i,{\bf p}_j\}=0, \{{\bf n}_i,{\bf n}_j\}=0. 
\end{equation}

To see the connection with Lie group representation, let us consider $W=V\oplus V'$ where $V$ and $V'$ are isotropic subspaces, with basis $(e_j)_{1\leq j\leq 2N}$, for the quadratic form $Q=\begin{pmatrix}
                                                                                                0 &I_N\\
                                                                                                I_N & 0
                                                                                               \end{pmatrix}$
and let us denote by $Cl(W,Q)$ the corresponding Clifford algebra \cite{FH}. Thus                                                                                               
$\mathcal{F}$ is a $Cl(W,Q)$ module 
\begin{equation} w=x_ie_i+y_j e_{N+j}\mapsto \sqrt{2} (x_i {\bf p}_i+y_j{\bf n}_j) \in End(\mathcal{F}).\end{equation}                                                                                              
It follows that $\mathcal{F}_+$ and $\mathcal{F}_{-}$ are irreducible representations of the simple Lie group $\text{Spin}(2N)$, i.e. 
the spin group\footnote{The spin group Spin$(2N)$ corresponds to the simply connected double cover of SO$(2N)$ \cite{FH}.}. Those irreducible representations are known as spinor representations.

\begin{ex}[The box picture]
 Let $V=\CC^{2n}=\CC^2\otimes \CC^n$, i.e. a  single particle can be in two different modes (${\up}$ or ${\down}$) and 
  $n$ different locations. We denote by 
 ${\bf p}_{1},\dots, {\bf p}_n, {\bf p}_{\overline{1}},\dots, {\bf p}_{\overline{n}}$ the corresponding creation operators where ${\bf p}_i$ creates an ${\up}$-particle in the $i$-th location 
 and ${\bf p}_{\overline{i}}$ creates a ${\down}$-particle in the $i$-th location.
 One can give a box picture representation of the embedding of $n$ qubits in the Hilbert space $\mathcal{F}=\mathcal{F}_+\oplus \mathcal{F}_{-}$.
 With the chirality decomposition $\mathcal{F}=\mathcal{F}_+\oplus \mathcal{F}_{-}$ one  gets two different ways of embedding $n$ qubits, Fig. \ref{spinor1} and Fig. \ref{spinor2}.

 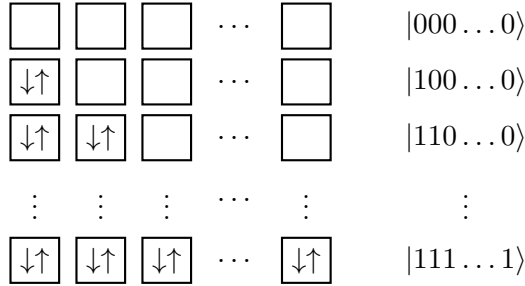
\begin{figure}[!h]
\begin{center}
\begin{tikzpicture}[scale=3]
    \tikzstyle{ann} = [draw=none,fill=none]
    \matrix[nodes={draw,  thick, fill=white!20},
        row sep=0.1cm,column sep=0.2cm] {
     \ee;&\ee; &\ee; &\node[ann] {$\dots$};  &\ee       ; &\node[ann] {~~~~~$|000\dots0\rangle$}; \\
      \du;&\ee; &\ee; &\node[ann] {$\dots$};  &\ee     ;   &\node[ann] {~~~~~$|100\dots0\rangle$};  \\
       \du;&\du; &\ee; &\node[ann] {$\dots$};  &\ee      ;  &\node[ann] {~~~~~$|110\dots0\rangle$};   \\
       \node[ann] {$\vdots$}; &\node[ann] {$\vdots$}; &\node[ann] {$\vdots$};&\node[ann] {$\dots$}; &\node[ann] {$\vdots$};&\node[ann] {~~~~~$\vdots$}; \\
        \du;&\du; &\du; &\node[ann] {$\dots$};  &\du   ;  &\node[ann] {~~~~~$|111\dots1\rangle$};  \\
    };
\end{tikzpicture}
\end{center}
\caption{Double occupancy embedding of the $n$-qubit Hilbert space ($2^n$ basis vectors) inside $\mathcal{F}_+$ ($n$ boxes and $N=2n$ single particle states).}\label{spinor1}
\end{figure}

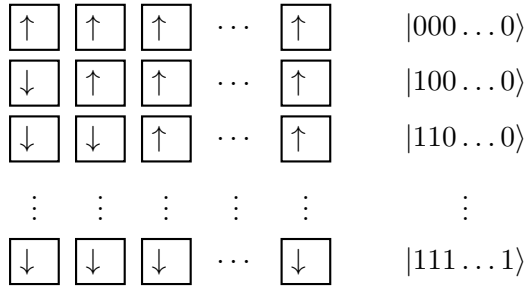
\begin{figure}[!h]
\begin{center}
\begin{tikzpicture}[scale=3]
    \tikzstyle{ann} = [draw=none,fill=none]
    \matrix[nodes={draw,  thick, fill=white!20},
        row sep=0.1cm,column sep=0.2cm] {
   &\ue;&\ue;&\ue; &\node[ann] {$\dots$};  &\ue    ; &\node[ann] {~~~~~$|000\dots0\rangle$};  \\
     &\de;&\ue;&\ue; &\node[ann] {$\dots$};  &\ue    ;  &\node[ann] {~~~~~$|100\dots0\rangle$};   \\
       &\de;&\de;&\ue; &\node[ann] {$\dots$};  &\ue    ;  &\node[ann] {~~~~~$|110\dots0\rangle$};  \\
  &\node[ann] {$\vdots$}; &\node[ann] {$\vdots$}; &\node[ann] {$\vdots$}; &\node[ann] {$\vdots$};&\node[ann] {$\vdots$};&\node[ann] {~~~~~$\vdots$};\\
       &\de;&\de;&\de; &\node[ann] {$\dots$};  &\de    ;  &\node[ann] {~~~~~$|111\dots1\rangle$};   \\
    };

\end{tikzpicture}
\end{center}
\caption{Single occupancy embedding of the $n$-qubit Hilbert space ($2^n$ basis vectors) inside $\mathcal{F}_{+}$ (for $n=2k$ boxes and $N=2n$ single particle states) or $\mathcal{F}_{-}$ (for $n=2k+1$ boxes and $N=2n$ single particle states).}\label{spinor2}
\end{figure}
\end{ex}

If we consider quantum information processing involving $n$ bosonic qubits, $n$ qubits or $n$ fermions with $2n$ modes, all systems 
can be naturally  embedded in the fermionic Fock space with $N=2n$ modes and the restriction of the action 
of the $\text{Spin}(2N)=\text{Spin}(4n)$ group to those sub-Hilbert-spaces boils down to their natural SLOCC group as shown in Table \ref{embed}.
In this sense the Spin group can be regarded as a natural generalization  of the SLOCC group.


\begin{table}[!h]
\begin{center}
 \begin{adjustbox}{max width=\textwidth}
 \begin{tabular}{l|lllllll}
Lie algebra&  $\mathfrak{s}\mathfrak{l}_2$                    & $\subset$ & $\mathfrak{s}\mathfrak{l}_2+\dots+\mathfrak{s}\mathfrak{l}_2$ & $\subset $& $\mathfrak{s}\mathfrak{l}_{2n}$ & $\subset$ &$\mathfrak{s}\mathfrak{o}_{4n}$\\
\hline
Lie group & $\SL_2(\CC)$         & $\subset$                   & $\SL_2(\CC)\times\dots\times \SL_2(\CC)$ &$\subset$     & $\SL_{2n}(\CC)$ &$\subset$  & $\text{Spin}(4n)$\\
\hline
Representation &  $\PP(\text{Sym}^n(\CC^2))$&$\hookrightarrow$  &$\PP(\CC^2\otimes \dots\otimes \CC^2)$ & $\hookrightarrow$& $\PP(\bigwedge ^n \CC^{2n})$ &$\hookrightarrow $&$\PP(\Delta_{4n})$\\
\hline
Highest weight orbit & $v_n(\PP^1)$ & $\subset$ & $\PP^1\times\dots\times \PP^1$ & $\subset$& $G(n,2n)$& $\subset$ & $\SS_{2n}$
 \end{tabular}
 \end{adjustbox}
 \caption{Embedding of $n$-bosonic qubit, $n$-qubit, $n$ fermions with $2n$ single particle states into fermionic Fock space with $2N=4n$ modes.}\label{embed}
 \end{center}
\end{table}
Let us denote by $\Delta_{4n}$ the irreducible representations $\mathcal{F}_{\pm}$, the algebraic variety $\SS_{2n}\subset \PP(\Delta_{4n})$ corresponding 
to the highest weight orbit of Spin$(4n)$
is called the spinor variety and generalizes the set of separable states.
Table \ref{embed} indicates that the classification of spinors could be considered as the general framework to study the entanglement 
classification of  pure quantum systems.  The embedding of qubits into fermionic systems (with a fixed number of particles) was used  
in \cite{CDG} to answer the question of SLOCC equivalence in the four-qubit case. In \cite{LH} we used the embedding within 
the fermionic Fock space to recover the polynomial invariants of the four-qubit case from the invariants of the spinor representation.

\subsection{From sequence  of simple Lie algebras to the classification of tripartite quantum systems with similar classes of entanglement}
Let us go back to the three qubits classification and the $\ket{W_3}$ and $\ket{GHZ_3}$ states. After the paper of D\"ur, Vidal and Cirac \cite{Dur} 
other papers were published in the quantum information literature describing other quantum systems featuring only
two 
types of genuine entangled states, similar to the $\ket{W_3}$- and $\ket{GHZ_3}$-states.

In  \cite{HL} we showed how all those similar classifications correspond to a sequence of varieties
 studied from representation theory and algebraic geometry in connection with the Freudenthal magic square \cite{LM2}. 
 Consider a Lie group $G$ acting by its adjoint action on its Lie algebra $\mathfrak{g}$. The adjoint variety $X_G\subset \PP(\mathfrak{g})$ 
 is the highest weight orbit for the adjoint action.  Take any point $x\in X_G$ and let us consider the set of all lines
 of $X_G$ passing through $x$ (these lines are tangent to $X_G$). This set of lines is a smooth homogeneous variety $Y\subset \PP(T_x X_G)$, called the subadjoint variety of $X_G$.
 Consider the sequence of Lie algebras 
 \begin{equation}
  \mathfrak{g}_2\subset \mathfrak{s}\mathfrak{o}_8\subset \mathfrak{f}_4\subset \mathfrak{e}_6\subset \mathfrak{e}_7.
 \end{equation}
This sequence gives rise to a series of subadjoint varieties called the {\em subexceptional series}. In \cite{LM2} this sequence is obtained as the third row of the geometric version of 
the Freudenthal's magic square.

To see how the subexceptional series is connected to the different classifications of \cite{brody2,LV,Sarosi,BDDER,DF,Levay1}
let us ask the following question: What do the Hilbert spaces $\mathcal{H}$ and the corresponding SLOCC groups $G$ look like such that 
the only genuine entanglement types are $\ket{W}$ and $\ket{GHZ}$ ?

If we assume that $G$ is a Lie group and $\mathcal{H}$ an irreducible representation such that the only two types of genuine entangled states are 
$\ket{W}$ and $\ket{GHZ}$ then one knows from Section \ref{tripartitesec} that the secant variety of the variety of separable states should fill the 
ambient space and be of the expected dimension.
 Because the secant variety is an orbit, this orbit is dense by our assumption and, therefore, the ring of SLOCC invariant polynomials 
should be generated by at most one element. But one also knows, under our assumption and by Zak's theorem, that in this case the tangential variety, i.e. the $\ket{W}$-orbit, is a codimension-one orbit in the ambient space. 
Thus the ring of $G$-invariant polynomials for  the representation $\mathcal{H}$ should be generated by a unique polynomial.
The classification of such representations was given in the 70's by Kac, Popov and Vinberg \cite{KPV}. From this classification 
one just needs to keep the representation where the dimension of the secant variety of the highest weight orbit is 
of the expected dimension. This leads naturally to the sequence of subexceptional varieties as given in Table \ref{subadjoint}.

\begin{table}[!h]
\begin{center}
\begin{adjustbox}{max width=\textwidth}
\begin{tabular}{c|c|c|c|c}
\hline
$\mathcal{H}$ & $\SLOCC$ & {QIT interpretation} & $X_{\Sep}\subset \PP(\mathcal{H})$ &  $\mathfrak{g}$\\
\hline
$Sym^3(\CC^2)$ & $\SL_2(\CC)$ & {Three bosonic qubit} \cite{brody2,VL2}& $v_3(\PP^1) \subset \PP^3$ & $\mathfrak{g}_2$\\
               &           &   (2007)                &                             &\\
\hline 
$\CC^2\otimes \CC^2\otimes \CC^2$ & $\SL_2(\CC)\times \SL_2(\CC)\times \SL_2(\CC)$ &  {Three qubit} \cite{Dur}&$\PP^1\times\PP^1\times \PP^1\subset\PP^7$ & $\mathfrak{so}_8$\\
                                   &                                             &  (2001)  &  &\\
\hline 
$\bigwedge ^{\langle 3\rangle} \CC^6$ & $\text{Sp}_6(\CC)$ & {Three fermions with} & $LG(3,6)\subset \PP^{13}$ & $\mathfrak{f}_4$\\
                                     &          &  {with }$6${ single particles state}    &      &       \\
                                     &          &   {with a symplectic condition}              &               & \\
                                     \hline
$\bigwedge^3\CC^6$ & $\SL_6(\CC)$ & {Three fermions with } & $G(3,6)\subset \PP^{19}$& $\mathfrak{e}_6$\\
                 &            & {with }$6${ single particles state} \cite{LV}& & \\
                 &            & (2008)                                      & &\\
                 \hline
   $\Delta_{12}$    & $\text{Spin}(12)$ & {Particles in Fermionic} &$\mathbb{S}_6\subset \PP^{31}$& $\mathfrak{e}_7$  \\
             &                & {Fock space}  \cite{Sarosi}      &  & \\
             &                &  (2014)                        && \\
 \hline
   $V_{56}$ &$ E_7$  & {Three partite entanglement} & $E_7/P_1 \subset \PP^{55}$&  $\mathfrak{e}_8$ \\
        &         & {of seven qubit}  \cite{DF,Levay1}         &   &\\
       &          &  (2007)                               &    & \\
       \hline
       &           & &  {\bf Freudenthal subexceptionnal series}&  \\   
\hline
\end{tabular}
\end{adjustbox}
\caption{The sequence of subexceptional varieties and the corresponding tripartite systems.}\label{subadjoint}
\end{center}
\end{table}
\begin{remark}\rm
 The relation between the Freudenthal magic square and the tripartite entanglement  was already pointed out in \cite{BDDER,VL2}. Other 
 subadjoint 
 varieties for the Lie algebra $\mathfrak{s}\mathfrak{o}_{2n}$, $n\neq 4$, not included in the subexceptional series also 
 share the same orbit structure. The physical interpretation of those systems is clear for $n=3,5,6$ \cite{HL,VL2}, but rather obscure in the general case $n\geq 7$.
\end{remark}

\begin{remark}\rm
This sequence of systems can also be considered from the dual picture by looking for generalization of the Cayley hyperdeterminant (the dual equation of $X=\PP^1\times\PP^1\times\PP^1$).
In \cite{LM2} it was also shown that all dual equations for the subexceptional series can be uniformly described. The tripartite entanglement of seven qubits \cite{DF},
under constrains given by the Fano plane, also started with a generalization of Cayley's quartic hyperdeterminant in relation 
with black-hole-entropy formulas in the context of the black-hole/qubit correspondence \cite{BDL}. 
\end{remark}

\section{The Geometry of Contextuality}\label{context}
In this second part of the paper I discuss the finite geometry behind operator-based proofs of contextuality. Starting from the geometric description of 
the $N$-qubit Pauli group, I recall how the concept of Veldkamp geometry associated to a point line configuration recently leaded us to recognize weight diagrams of 
simple Lie algebras in some specific arrangement of hyperplanes of the three-qubit Pauli group. 
\subsection{Observable-based proofs of contextuality}
As explained in the introduction, operator-based proofs of the Kochen-Specker (KS) Theorem correspond to configurations 
of mutli-Pauli observables such that the operators on the same context (line) are mutually commuting and such that the product
of the operators 
gives $\pm I$, with an odd number of negative contexts.

\begin{figure}[!h]
 \begin{center}
  \includegraphics[width=6.5cm]{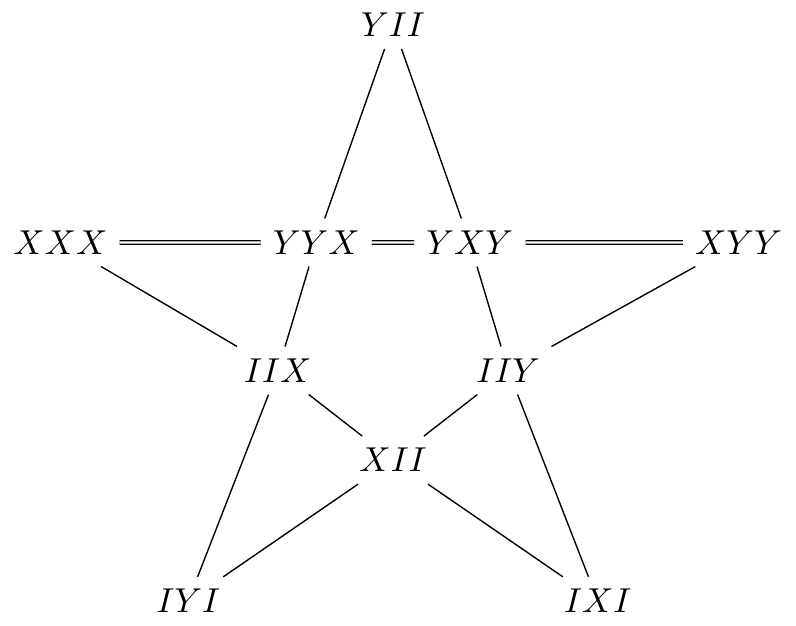}
\caption{The Mermin pentagram: a configuration of 10 three-qubit operators proving KS Theorem. Operators on a line are mutually commuting and the doubled line corresponds to the context where the product gives $-I_8$.}\label{penta}
\end{center}
  \end{figure}

The Mermin-Peres square presented in the introduction  is the first operator/observable-based proof of the KS Theorem. In \cite{mermin}, Mermin 
also proposed another proof involving three qubit Pauli operators and known as the Mermin pentagram (Fig. \ref{penta}).

 The Mermin-Peres square and the Mermin pentagram are the smallest configurations, in terms of number of 
 contexts and number of operators,  
 providing observable based proofs of contextuality \cite{HS}. Other proofs of the KS Theorem based on observable configurations 
 have been proposed by  Waegel and Aravind \cite{WA1,WA2} or Planat and Saniga \cite{PS,SP}.
 In terms of quantum processing, the <<magic>> configurations have been investigated under the scope of non-local games. For each magic configuration
 one can define a game where cooperative players can win with certainty using a quantum strategy.
 Let us look at  
 the magic square of 
 Fig. \ref{square} and consider the following game involving two players Alice and Bob, and a referee Charlie. As usual, Alice and Bob may define a strategy in advance but cannot communicate once the game starts:
 \begin{enumerate}
  \item Charlie picks a number $r\in\{1,2,3\}$ for a row and $c\in\{1,2,3\}$ for a column and sends $r$ to Alice and $c$ to Bob.
  \item Both Alice and Bob send back to the referee a triplet of $\pm 1$ such that the number of $-1$ is odd for Alice and even for Bob.
  \item Alice and Bob win the game if the number in position $c$ of Alice triplet matches with the number in position $r$ for Bob's triplet (and of course the triplets of Alice and Bob satisfy the parity condition of the previous step).
 \end{enumerate}
Such type of game is called a binary constrain game \cite{cleve}. If Alice and Bob share a specific four-qubit entangled state (a product of two $\ket{EPR}$-like states) they can win that game with certainty, while it is easy to prove that 
there is no such classical strategy. In \cite{Arkhipov}, Arkhipov gave a graph-theoretic characterization of 
magic configurations in  terms of 
planarity of the dual configuration.

A natural question to ask is to find all possible different realizations of a given magic configuration.  For instance 
one can ask how many two-qubit KS proof similar  to the Mermin-Peres square can be built, or how many Mermin pentagrams can we 
obtain with three-qubit Pauli operators ? As we will explain now this can be answered by looking at the geometry of 
the space of $N$-qubit Pauli operators.

\begin{remark}\rm
 Originally, the first proof of KS was not given in terms of configurations of multiqubit Pauli-operators, but by considering projection operators on some specific basis of the 
 three-dimensional Hilbert space. Kochen and Specker found a set of 117 operators and proved the impossibility to assign a deterministic value $\pm 1$ to each of them by using a coloring argument 
 on the corresponding basis vectors. Several simplification of this original proof were proposed in the literature. For 
 instance, one can reduce to $18$ the number of vectors needed 
 to express the KS Theorem in terms of projectors \cite{cabello}.
\end{remark}

\subsection{The Symplectic Polar space of rank $N$ and the $N$-qubit Pauli group}
To understand where these magic configurations live, we now start to describe geometrically the generalized $N$-qubit Pauli group, i.e. the group of Pauli operators 
acting on $N$-qubit systems.
The following construction is due to M. Saniga and M. Planat \cite{SP,HOS,Thas} and has been employed in the past $10$ years 
to provide a finite geometric insight starting from the commutation relations of Pauli observables up to the black-hole-entropy formulas \cite{BDL,LPS,LSVP}.

Let us consider the subgroup $P_N$ of $GL(2^N,\C)$ generated by the tensor products of Pauli matrices,
\begin{equation} A_1\otimes A_2\otimes\dots\otimes A_N\equiv A_1A_2\dots A_N,\end{equation}
with $A_i\in \{\pm I, \pm i I,\pm X,\pm i X, \pm Y, \pm iY, \pm Z, \pm i Z\}$. The center of $P_N$ is $\mathcal{C}(P_N)=\{\pm I, \pm i I\}$ and $V_N=P_N/\mathcal{C}(P_N)$ is an abelian group.

To any class $\overline{\mathcal{O}}\in V_N$, there corresponds a unique element in $\Z_2 ^{2N}$. More precisely,  for any $\mathcal{O}\in P_N$
 we have $\mathcal{O}=sZ^{\mu_1}X^{\nu_1}\otimes \dots \otimes Z^{\mu_N} X^{\nu_N}$ with $s\in \{\pm 1,\pm i\}$ and $(\mu_1,\nu_1,\dots, \mu_N,\nu_N)\in \Z_2 ^{2N}$.
 Thus $V_N$ is a $2N$ dimensional vector space over 
 $\Z_2$ and we can associate to any non-trivial observable $\mathcal{O}\in P_N \setminus I^N$ a unique point 
 in the projective space $\PP^{2N-1} _{2}=\PP(\ZZ_2^{2N})$.
\begin{equation}
 \begin{array}{llll}
  \pi: &P_N\setminus I_N & \to & \PP_2 ^{2N-1} \\
  & \mathcal{O}=sZ^{\mu_1}X^{\nu_1}\otimes \dots \otimes Z^{\mu_N} X^{\nu_N} & \mapsto & [\mu_1:\nu_1:\dots: \mu_N:\nu_N].
 \end{array}
\end{equation}
Because $V_N$ is a vector space over $\Z_2$, the lines of  $\PP_2 ^{2N-1}$ are made of triplet of points $(\alpha,\beta,\gamma)$ such that $\gamma=\alpha+\beta$.
The corresponding (class) of observables
$\overline{\mathcal{O}_\alpha}$, $\overline{\mathcal{O}_\beta}$ and $\overline{\mathcal{O}_\gamma}$ satisfy 
$\overline{\mathcal{O}_\alpha}.\overline{\mathcal{O}_\beta}=\overline{\mathcal{O}_\gamma}$ ($.$ denotes the ordinary product of operators).
 
 \begin{ex}
 For single qubit we have $\pi(X)=[0:1]$, $\pi(Y)=[1:1]$ and $\pi(Z)=[1:0]$. The projective space $\PP^1_2$ is the projective line $(X,Y,Z)$ (the projection $\pi$ will be omitted).
\end{ex}

However, the correspondence between non-trivial operators of $P_N$ and points in $\PP_2^{2N-1}$ does not say anything about the 
commutation  relations between the operators. To see geometrically these commutation relations, one needs to introduce an extra structure.
Let $\mathcal{O},\mathcal{O}'\in P_N$ such that $\mathcal{O}=sZ^{\mu_1}X^{\nu_1}\otimes \dots\otimes Z^{\mu_N} X^{\nu_N}$ and $\mathcal{O}'=s'Z^{\mu_1'}X^{\nu_1'}\otimes \dots \otimes Z^{\mu_N'} X^{\nu_N'}$ 
with  $s,s'\in \{\pm 1,\pm i\}$ and $\mu_i,\nu_i,\mu_i',\nu_i'\in \Z_2$.

Then, we have \begin{equation} \mathcal{O}.\mathcal{O}'=(ss'(-1)^{\sum_{j=1} ^N \mu_j'\nu_j},\mu_1+\nu_1',\dots,\mu_N+\nu_N'),\end{equation}
and the two elements $\mathcal{O}$ and $\mathcal{O}'$ of $P_N$ commute, if and only, if 
\begin{equation} \sum_{j=1} ^N (\mu_j\nu'_j+\mu_j'\nu_j)=0.\end{equation}
Let us add to $V_N$  the symplectic form \begin{equation}\langle \overline{\mathcal{O}}, \overline{\mathcal{O}'}\rangle= \sum_{j=1} ^N (\mu_j\nu'_j+\mu_j'\nu_j),\end{equation}
and let us denote by $\mathcal{W}(2N-1,2)$ the symplectic polar space of rank $N$ i.e. the set of totally isotropic
subspaces of $(\PP^{2N-1}_2,\langle,\rangle)$.
The symplectic polar space $\mathcal{W}(2N-1,2)$ encodes the commutation relations of $P_N\setminus I_N$. The points of $\mathcal{W}(2N-1,2)$ correspond to non trivial operators 
of $P_N$ and 
the subspaces of $\mathcal{W}(2N-1,2)$ correspond to $\PP(\mathcal{S}/\mathcal{C}(P_N))$, where $\mathcal{S}$ is a set of mutually commuting elements of $P_N$. 
\subsection{Geometry of hyperplanes: Veldkamp space of a point-line geometry}
The points and lines of  $\mathcal{W}(2N-1,2)$ define an incidence structure, i.e. a point-line geometry $\mathcal{G}=(\mathcal{P},\mathcal{L},\mathcal{I})$ where $\mathcal{P}$ 
are the points of $\mathcal{W}(2N-1,2)$,  
$\mathcal{L}$ are the lines and  $\mathcal{I}\subset \mathcal{P}\times\mathcal{L}$  corresponds to the incidence relation. 
 I now introduce some geometric notions for point-line incidence structures.
\begin{definition}
 Let $\mathcal{G}=(\mathcal{P},\mathcal{L},\mathcal{I})$ be a point-line incidence structure.  A hyperplane $H$ of $\mathcal{G}$ is a subset of 
 $\mathcal{P}$
 such that a line of $\mathcal{L}$ is either contained in $H$, or has a unique intersection with $H$.
\end{definition}

\begin{ex}
 Let us consider a $3\times3$ grid  with $3$ points per line, also known as $GQ(2,1)$. This geometry has $15$ hyperplanes splitting in 
 two different types: the perp sets (the unions of two <<perpendicular>> lines) and the ovoids (hyperplanes that contain no lines).
\begin{figure}[!h]
 \begin{center}
 \includegraphics[width=7cm]{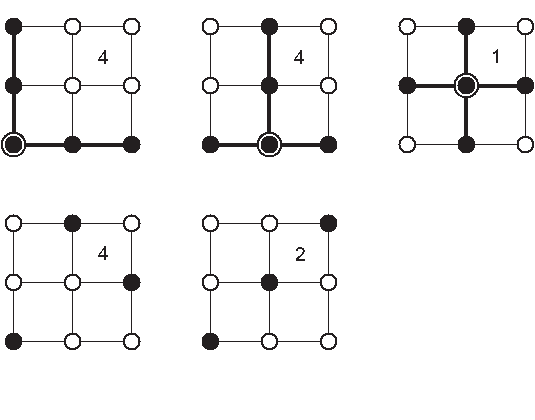}
  \caption{Pictural representation of the $15$ hyperplanes of the grid $GQ(2,1)$. $9$ hyperplanes are of type perp and $6$ of them are of type ovoid.}
 \end{center}
\end{figure}
 \end{ex}

The notion of geometric hyperplanes leads to the notion of Veldkamp space as introduced in \cite{SPPH}.
 \begin{definition}
  Let $\mathcal{G}=(\mathcal{P},\mathcal{L},\mathcal{I})$ be a point-line geometry. The Veldkamp space of $\mathcal{G}$, denoted by $\mathcal{V}(\mathcal{G})$, if it exists, is a point-line geometry
  such that
  \begin{itemize}
   \item the points of $\mathcal{V}(\mathcal{G})$ are geometric hyperplanes of $\mathcal{G}$,
   \item given two points $H_1$ and $H_2$ of $\mathcal{V}(\mathcal{G})$, the Veldkamp line defined by $H_1$ and $H_2$ is the set of hyperplanes of $\mathcal{G}$ such that 
   $H_1\cap H=H_2\cap H$ or $H=H_i, i=1,2$.
  \end{itemize}
 \end{definition}

\begin{figure}[!h]
 \begin{center}
\includegraphics[width=7cm]{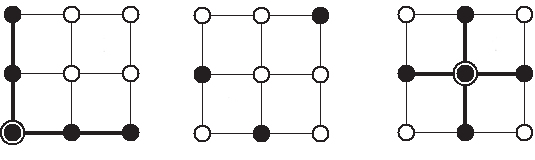}
  \caption{An example of Veldkamp line of the grid, i.e. a line of $\mathcal{V}(GQ(2,1))$. The three hyperplanes share two by two the same intersection (and no other hyperplane of $GQ(2,1)$ does). Taking two of those three hyperplanes, the thrid one is obtained by considering the complement of the symmetric difference (see Eq (\ref{symmdiff}) below).}\label{veldkampgrid}
 \end{center}
\end{figure}

Fig. \ref{veldkampgrid} furnishes an example of a Veldkamp line in $\mathcal{V}(GQ(2,1))$. It is not to difficult to show that the 
hyperplanes of $GQ(2,1)$ accommodate the $15$ points and $35$ lines of $\PP_2 ^3$, i.e. $\mathcal{V}(GQ(2,1))=\PP^3 _2$.

\begin{remark}\rm
 The notion of Veldkamp space of finite point-line incidence structures has been employed to study orbits in $\PP^{2^N-1}_2$ under the action  
 of $\SL_2(\F_2)\times \dots\times \SL_2(\F_2)$. For $N=4$, it was possible to obtain a computer free proof of the  
 classification of the  $2\times 2\times 2\times 2$ tensors over $\F_2$
 by classifying the hyperplanes of a specific configuration. More precisely it was shown in \cite{SHHPP} that 
 the Veldkamp space of the finite Segre varieties of type $S_N=\underbrace{\PP^1 _{2}\times\dots\times \PP^1_2}_{N \text{ times}}$ is 
 the projective space 
 $\PP_2 ^{2^N-1}$ and that the different types of hyperplanes of $S_N$ are in bijection with the $\SL_2(\F_2)\times \dots\times \SL_2(\F_2)$-orbits 
 of $\PP^{2^N-1}_2$. 
\end{remark} 
\subsection{The finite geometry of the $2$-qubit and $3$-qubit Pauli groups and the hyperplanes of $\mathcal{W}(2N-1,2)$}
  
We now describe in detail $\mathcal{W}(3,2)$ and $\mathcal{W}(5,2)$, the symplectic polar spaces encoding the commutation 
relations of the $2$ and $3$-qubit Pauli groups and their Veldkamp spaces.

\begin{figure}[!h]
  \begin{center}
 \includegraphics[width=7cm]{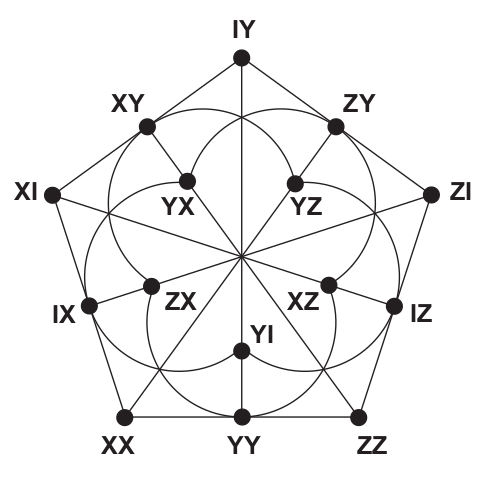}
 \caption{The labeling of the doily, i.e. the symplectic polar space $\mathcal{W}(3,2)$, by Pauli operators. The doily is a $15_3$-configuration ($15$ points, $15$ lines, $3$ points per line and $3$ lines through each point) which is a generalized quadrangle (i.e. is triangle free). It is the unique $15_3$-configuration that is triangle free among $245342$ ones. The doily encodes the commutation relations of the two-qubit Pauli group.}\label{doily}
\end{center}
 \end{figure}

 The symplectic polar space $\mathcal{W}(3,2)$ consists of all $15$ points of $\PP^3 _2$ but only the $15$ 
 isotropic lines are kept. This gives a point-line configuration description of $\mathcal{W}(3,2)$,  Fig. \ref{doily}, known as the doily \cite{HOS}.

  The doily is also known as the generalized quadrangle\footnote{A point-line incidence structure is 
  called a generalized quadrangle of type $(s,t)$, and denoted by  $GQ(s,t)$ 
   iff it is an incidence structure such that every point is on $t+1$ lines
  and every line contains $s+1$ points such that if $p\notin L, \exists!q\in L$ such that $p$ and $q$ are collinear.} $GQ(2,2)$. 
  In the following I will keep denoting by $\mathcal{W}(3,2)$ both the symplectic polar space and the associated point-line geometry $GQ(2,2)$.  
Looking at the doily (Fig. \ref{doily}) one can identify the  Mermin-Peres squares built with two-qubit 
Pauli operators as geometric hyperplanes of $\mathcal{W}(3,2)$. 
 
 \begin{figure}[!h]
 \begin{center}
  \includegraphics[width=7cm]{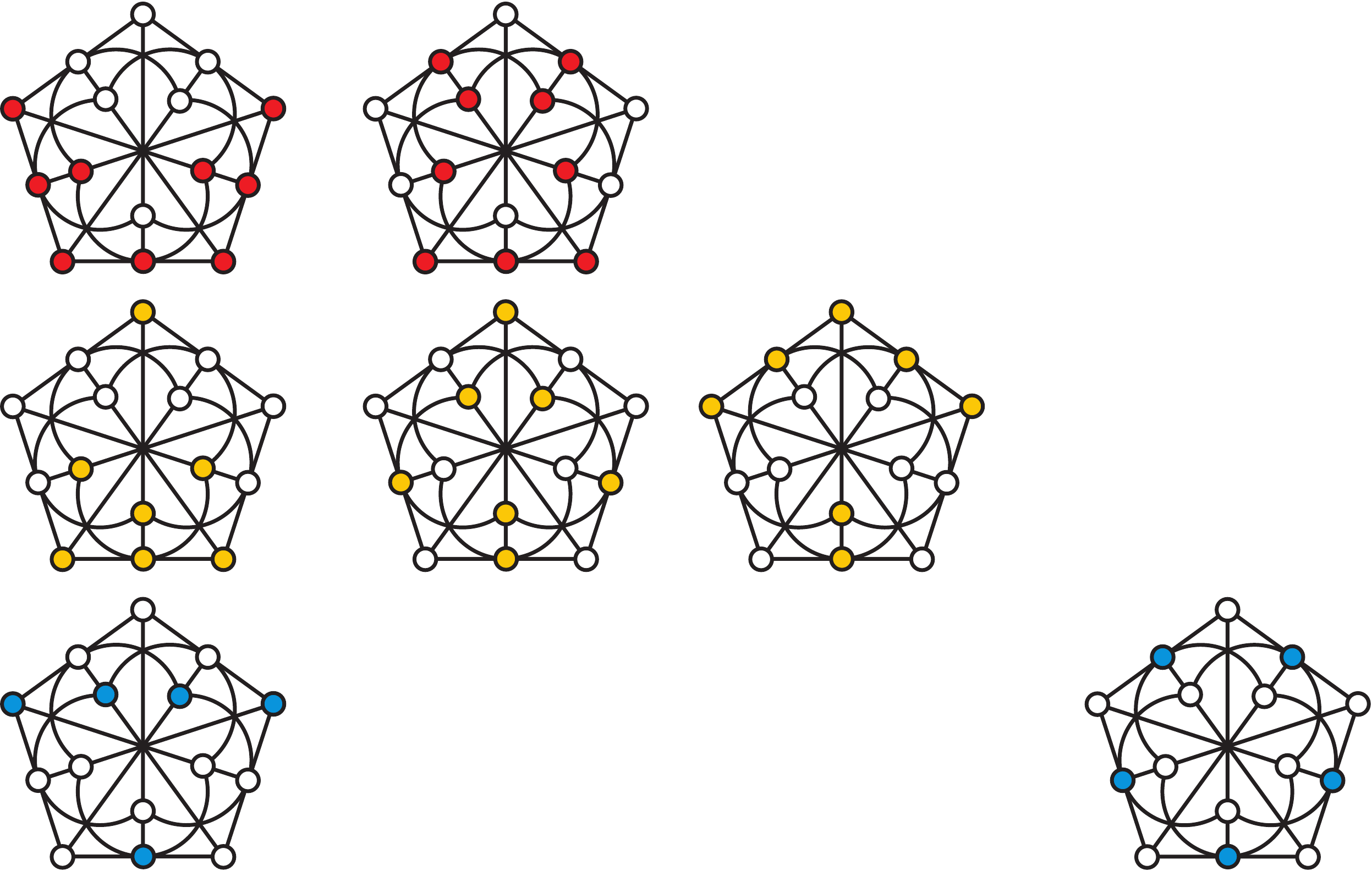}
  \caption{The three different types of hyperplanes of the doily \cite{SPPH}. In red hyperplanes corresponding to grids, $GQ(2,1)$, in yellow hyperplanes corresponding to perp-sets and in blue hyperplanes of type ovoids.}\label{hdoily}
 \end{center}
 \end{figure}
In fact,  three different types of hyperplanes can be found in the doily as shown in  Fig. \ref{hdoily}:
\begin{itemize}
\item The hyperplanes made of 9 points (red) correspond to grids $GQ(2,1)$ and it is easy to check 
 that  grids on the two-qubit Pauli group are always contextual configurations \cite{HS}, i.e. Mermin-Peres squares.
 Rotating by $\dfrac{2\pi}{5}$  one gets $10$ Mermin-Peres grids in the doily. 
 \item The second type of 
 hyperplanes (yellow ones) are called perp-sets (all lines of the hyperplane meet in one point) 
 and one sees from Fig. \ref{hdoily} that there are $15$ of such. 
 \item Finally the last type of hyperplanes of the 
 doily (blue) are line-free and such type of hyperplanes are called ovoids. The doily contains $6$ ovoids.
 \end{itemize}

\begin{figure}[!h]
  \begin{center}
  \includegraphics[width=6cm]{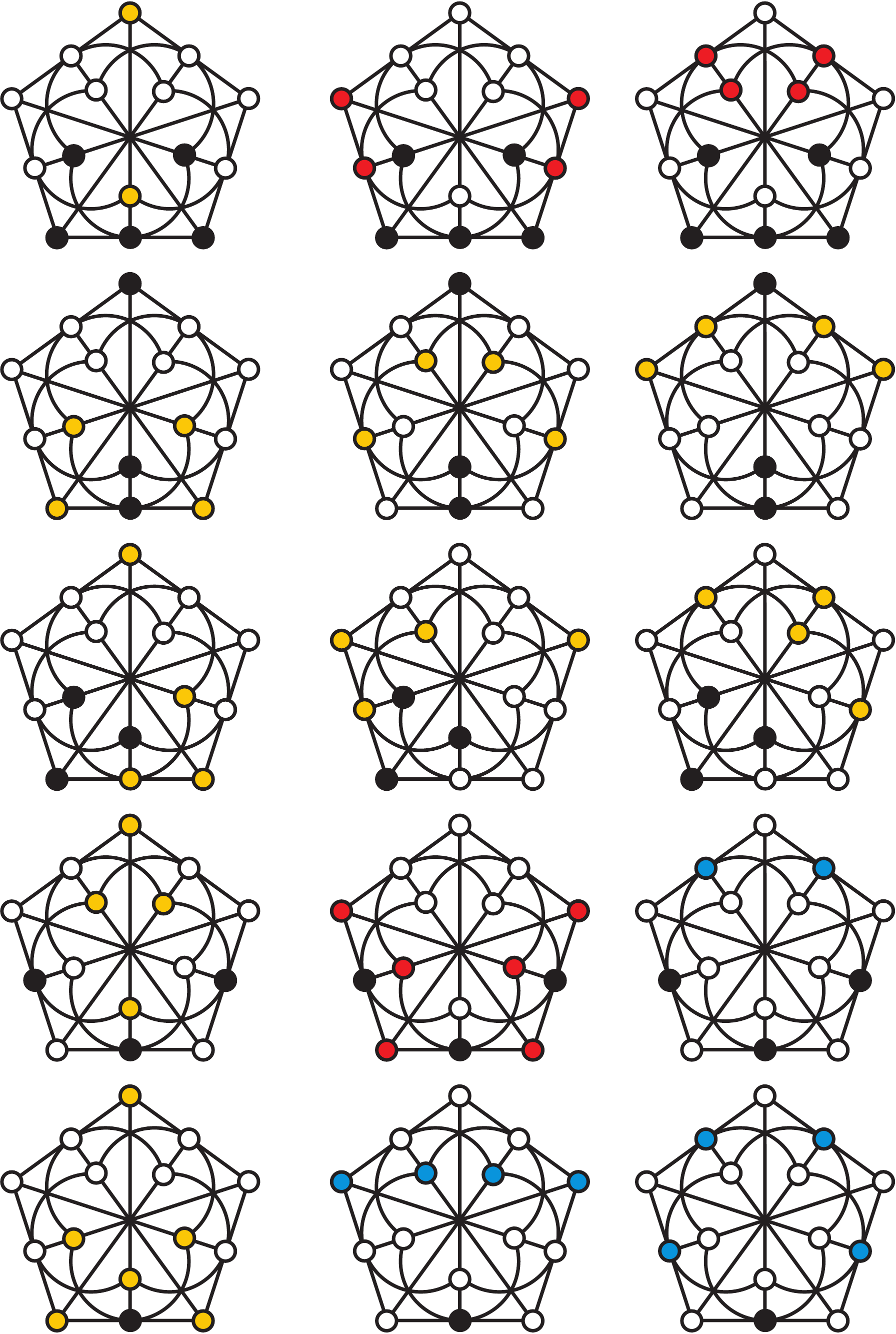}
  \caption{The $5$ types of Veldkamp lines of the doily \cite{SPPH}. For each line, the points collored in black correspond to the core of the Velkamp line. Note that two types of lines (the second and third) have the same composition (perp-perp-perp) and are distinguished by the core set, which is either composed of three noncolinear points or three points on a line. One can check that given any two hyperplanes on a line, the third one is  the complement of the symmetric 
  difference of the two, see Eq (\ref{symmdiff}).}\label{vldoily}
  \end{center}
 \end{figure}

 The geometry of $\mathcal{V}(GQ(2,2))$, the Veldkamp space of the doily, is described in full details in \cite{SPPH}.
 Fig. \ref{vldoily} illustrates the different types of Veldkamp lines that can be obtained from the hyperplanes of the doily.
 
 In particular, $\mathcal{V}(\mathcal{W}(3,2))$ comprises $31$ points splitting in three orbits and $155$ lines splitting in $5$ different types. One can show
 that $\mathcal{V}(GQ(2,2))\simeq \PP^4 _2$.
 

  
   


%
       The symplectic polar space $\mathcal{W}(5,2)$ contains $63$ points, $315$ lines, $135$ Fano planes.
       One can build $12096$ distinguished Mermin pentagrams from those $63$ points \cite{PSH,LPS}.

\begin{figure}[!h]
 \begin{center}
 \includegraphics[width=10cm]{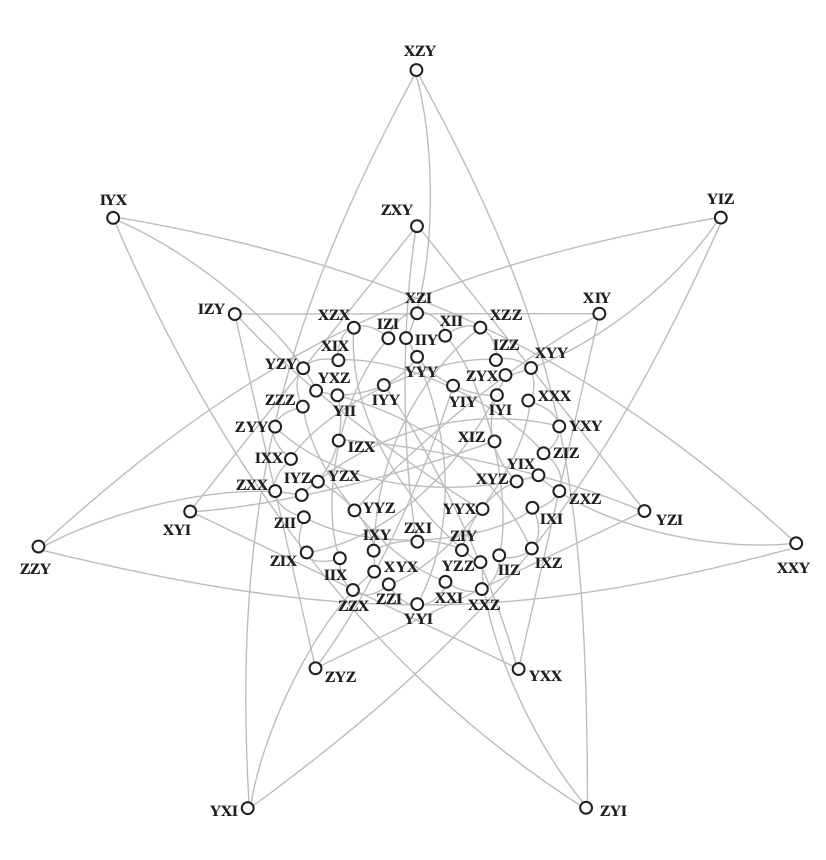}
 \caption{A 3-qubit Pauli group embedding into the split Cayley hexagon \cite{LSVP} in $\mathcal{W}(5,2)$. The split Cayley hexagon is a generalized polygon, it is a $63_3$ configuration that contains no ordinary pentagon.}\label{sch}
 \end{center}
\end{figure}

       In the case of the three-qubit Pauli group there is no generalized polygon which accommodates the full geometry 
       $\mathcal{W}(5,2)$.  However, there exists  an embedding in $\mathcal{W}(5,2)$ of the split-Cayley hexagon of order two, 
       which is a generalized hexagon of $63$ points and $63$ lines such that 
each line contains 3 points and each point belongs to $3$ lines. 
This split-Cayley hexagon accommodates the $63$ three-qubit operators of the three-qubit Pauli group 
 such that the lines of the configuration are totally isotropic lines (Fig. \ref{sch}).
 
The general structure of $\mathcal{V}(\mathcal{W}(2N-1,2))$ has been studied in details in \cite{VL} where 
the description of the geometric hyperplanes of $\mathcal{W}(2N-1,2)$ is explicitly given. First, let us mention that 
   for $\mathcal{G}=\mathcal{W}(2N-1,2)$ the  Veldkamp line defined by two hyperplanes $H_1$ and $H_2$ 
   is a  3-point line $(H_1,H_2,H_3)$ where $H_3$ is 
 given by the complement of the symmetric difference,

 \begin{equation}\label{symmdiff} 
 H_3=H_1\boxplus H_2=\overline{H_1\Delta H_2}.
 \end{equation}
To reproduce the description  of $\mathcal{V}(\mathcal{W}(2N-1,2)$ of  \cite{VL}, let us introduce the following quadratic form over $V_N$:
\begin{equation}
 Q_0(x)=\sum_{i=1}^N a_ib_i \text{ where } x=(a_1,b_1,\dots,a_N,b_N).
\end{equation}
An observable $\mathcal{O}$ is said to be symmetric if it contains an even number of $Y$'s or skew-symmetric if it contains an odd number of $Y$'s. 
In terms of the quadratic form $Q_0$, this leads to the conditions $Q_0(\overline{\mathcal{O}})=0$ or $Q_0(\overline{\mathcal{O}})=1$.

There are three types of geometric hyperplanes in $\mathcal{W}(2N-1,2)$:
 \begin{equation}
\text{Type 1: }  C_q=\{p \in \mathcal{W}(2N-1,2), \langle p, q\rangle=0\}.
 \end{equation}

 This set corresponds to the <<perp-set>> defined by $q$, i.e. in terms of operators, it is the set of elements commuting with $\overline{\mathcal{O}}_q$.
 
 To define  Type 2 and Type 3, let us introduce a family of quadratic forms on $V_N$ parametrized by the elements of $V_N$: $Q_q(p)=Q_0(p)+\langle q,p\rangle$.
 Depending on the nature of $\overline{\mathcal{O}}_q$ (symmetric or skew-symmetric) the quadratic form will be hyperbolic or elliptic.
 \begin{equation}
 \text{Type 2: } \text{for }\overline{\mathcal{O}}_q \text{ symmetric }H_q=\{p \in \mathcal{W}(2N-1,2), Q_q(p)=0\}\simeq Q^+(2N-1,2),
 \end{equation}
and 
 \begin{equation}
 \text{Type 3: } \text{for }\overline{\mathcal{O}}_q \text{ skew-symmetric }H_q=\{p \in \mathcal{W}(2N-1,2), Q_q(p)=0\}\simeq Q^-(2N-1,2),
 \end{equation}
 where $Q^{+}(2N-1,2)$ denotes a hyperbolic quadric\footnote{Up to a transformation of coordinates, this is a set of points  $x\in \PP^{2N-1}_2$ 
 satisfying the standard equation $x_1x_2+x_3x_4+\dots+x_{2N-1}x_{2N}=0$.} of $\mathcal{W}(2N-1,2)$, and  $Q^{-}(2N-1,2)$ denotes an elliptic quadric\footnote{Up to a transformation of coordinates this is 
 defined as points $x\in \PP^{2N-1}_2$ such that $f(x_1,x_1)+x_2x_3+\dots x_{2N-1}x_{2N}=0$.} of $\mathcal{W}(2N-1,2)$.
 
The set $H_q$ represents the set of observables  either symmetric and commuting with $\overline{\mathcal{O}}_q$ or skew-symmetric and anticommuting with $\overline{\mathcal{O}}_q$.

Moreover, the following equalities hold \begin{equation} C_p\boxplus C_q=C_{p+q}, H_p\boxplus H_q=C_{p+q} \text{ and }C_p\boxplus H_q=H_{p+q}.\end{equation}

This leads to five  different types of Veldkamp lines in $\mathcal{W}(2N-1,2)$ depending on the nature (symmetric or not) of the points $p$ and $q$ (we recover the $5$ different types of Veldkamp lines 
 illustrated in Fig \ref{vldoily}).

%

 \subsection{From commutation relations of the 3-qubit Pauli group to the weight diagrams of simple Lie algebras}
It was first pointed out in \cite{LS} that the Mermin pentagrams showing up in the three-qubit Pauli group 
can all be obtained from a <<double six>> configuration of such pentagrams living in  a Veldkamp line of type perp-hyperbolic-elliptic.
More precisely, taking the transitive action of the symplectic group $\text{Sp}(6,2)$ on $\mathcal{W}(5,2)$, one can recover all Mermin pentagrams from the 
$12$ pentagrams living in a specific subspace of the Veldkamp line $(H_{III},H_{YYY},C_{YYY})$.
According to the previous subsection one has the following description of the three hyperplanes $H_{III}, H_{YYY}$ and $C_{YYY}$ in terms of Pauli operators,
\begin{itemize}
 \item $C_{YYY}$ is the perp-set defined by the operator $YYY$, i.e. the points in $C_{YYY}$ correspond to operators commuting with $YYY$.
 \item $H_{III}$ is a hyperbolic quadric, i.e. is defined by $Q_0(x)=0$. In terms of operators it corresponds to the set of symmetric operators (i.e. containing an even number of $Y$).
 \item $H_{YYY}$ is an elliptic quadric, i.e. is defined by $Q_{YYY}(x)=0$. In terms of operators it corresponds to the set of symmetric operators commuting with $YYY$ or skew-symmetric ones anti-communting with $YYY$.
\end{itemize}

\begin{figure}[!h]
 \begin{center}
  \includegraphics[width=7cm]{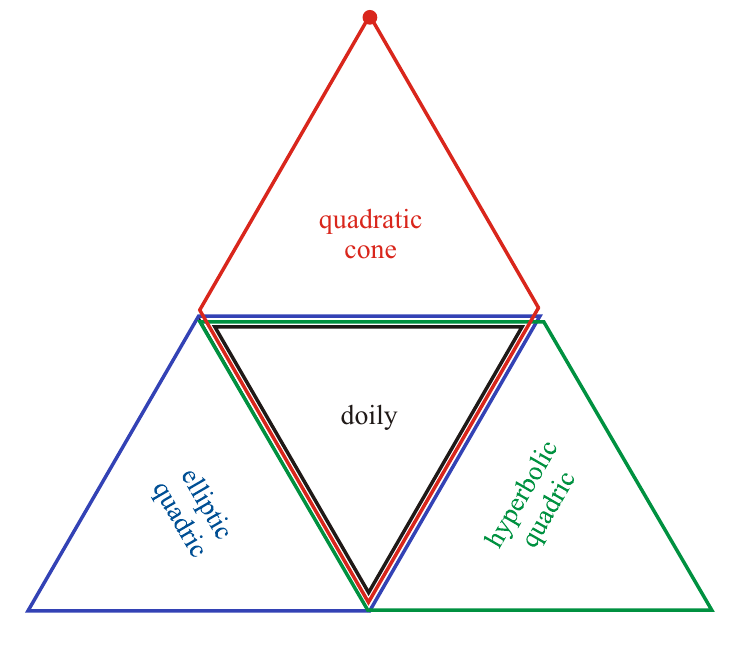}
  \caption{Schematic representation of the Veldkamp line $(H_{III},H_{YYY},C_{YYY})$. The core set of this $3$-qubit Veldkamp line is made of $15$ operators. The commutation relations among those $15$ operators define the incidence structure of a doily.}\label{veldkamline3qubit}
 \end{center}
 \end{figure}

\begin{figure}[!h]
 \begin{center}
 \includegraphics[width=9cm]{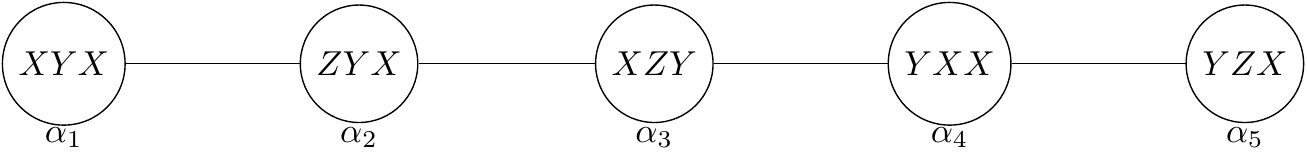}
 \caption{Labelling of the Dynkin diagram of type $A_5$ by $3$-qubit Pauli operators.}\label{roots}
\end{center}
 \end{figure}

\begin{figure}[!h]
\begin{center}
 \includegraphics[height=8cm]{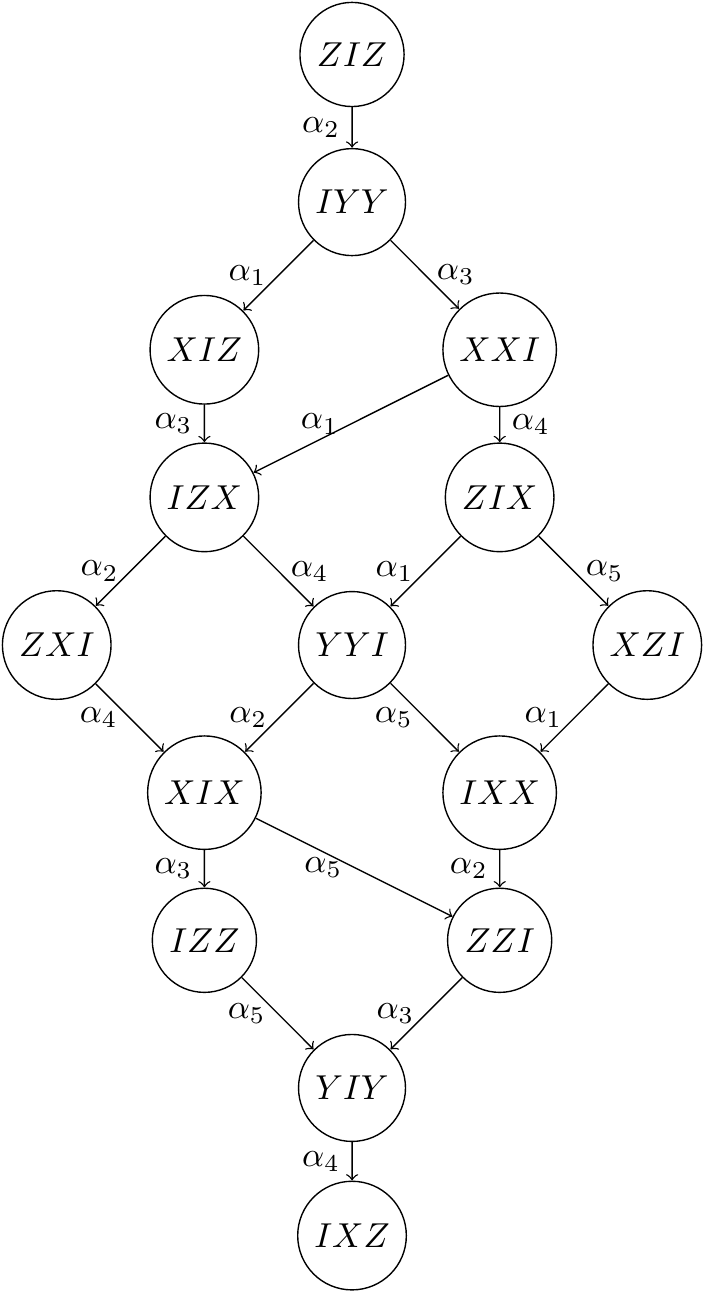}
 \caption{The weight diagram of the $15$-dimensional representation of $A_5$ in  terms 
 of $3$-qubit operators. The action by the roots $\alpha_1,\dots,\alpha_5$ of the Dynkin diagram is obtained by multiplying the weight by the $3$-qubit operator corresponding to the root (Figure \ref{roots}).}\label{A5diag}
\end{center}
 \end{figure}

 The core set of the Veldkamp line is the set of elements commuting with $YYY$ (they belong to $C_{YYY}$) and symmetric (they belong to $H_{III}$).
 An explicit list of those elements is given by:
 
\begin{equation}
 \begin{array}{cccccccc}
  YYI & YIY & IYY & ZZI & ZIZ & IZZ & XXI & XIX \\ 
  IXX &ZXI & ZIX & IZX &XZI & XIZ & IXZ. & \\
 \end{array}
\end{equation}
This set of observables forms a doily in $\mathcal{W}(5,2)$ (see Fig. \ref{veldkamline3qubit}) that encapsulates the weight diagram of the second fundamental
representation of $A_5$.
To see this connection with simple Lie algebras, let us associate to the roots 
$\alpha_1,\dots,\alpha_5$ of $A_5$ five skew-symmetric observables as given in Fig \ref{roots}.
The action of the roots by translation on the weight vectors \cite{FH} corresponds to multiplication in terms of operators. Now taking $ZIZ$ as the highest weight vector, 
then Fig. \ref{A5diag} reproduces the weight diagram
of the $15$-dimensional irreducible representation of $A_5$ built in terms of the $3$-qubit operators corresponding to the doily of Fig. \ref{veldkamline3qubit}.

\begin{figure}[!h]
 \begin{center}
  \includegraphics[width=5cm]{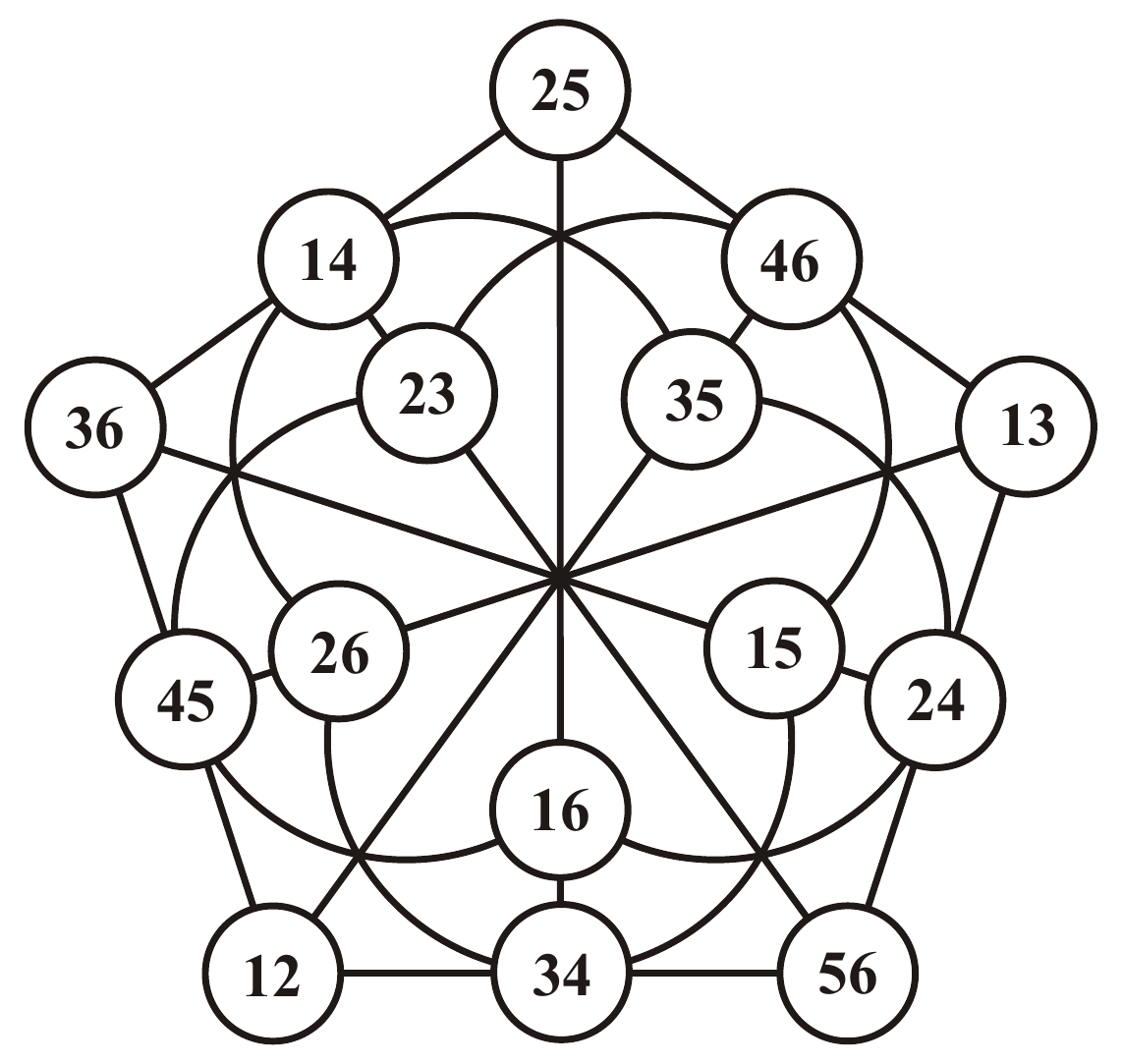}
  \caption{Labeling of the doily by doublets. Two doublets are colinear if they have no element in common (the third doublet on the line being the complement of the two doublets).}\label{doily2}
  \end{center}
 \end{figure}

This core set also encodes the Pfaffian of $6\times 6$ skew-symmetric matrices which is the invariant of the $15$-dimensional irreducible representation of $A_5$.
To see this, consider the observable $\Omega=\sum_{1\leq i< j\leq 6} a_{ij} \mathcal{O}_{ij}$, 
where $\mathcal{O}_{ij}$ is a three-qubit observable located at $(ij)$ (Fig \ref{doily2}).
Then the polynomial  $Tr(\Omega^3)$ is proportional to the Pfaffian, $Pf(A)$, where $A=(a_{ij})_{1\leq i <j \leq 6}$ is a skew symmetric matrix.
\begin{remark}\rm
A different choice of representatives of the root system of $A_5$ will generate a different weight diagram with the 
operators composing the doily, i.e. the choice of the representatives of the 
roots determines the highest weight vector.
 In \cite{LHS} we provided a labeling of the operators of the Veldkamp line in terms of a Clifford algebra. This has the double advantage to avoid a specific labeling but also 
 establish a connection of the full Veldkamp line
 with the $\text{Spin}(14)$ representation.
\end{remark}

Similarly, all hyperplanes of the Veldkamp line $(C_{YYY},H_{YYY},H_{III})$ can be analyzed this way, revealing connection with the $27$-dimensional 
irreducible representation of $E_6$ (elliptic quadric) or the $35$-dimensional irreducible representation of $A_6$ (hyperbolic quadric) as well as their corresponding invariants (Table \ref{inva}).
Subparts (triangles in Fig. \ref{veldkamline3qubit}) can be combined to get other irreducible representations like the $32$-dimensional irreducible 
representation of $\text{SO}(12)$ which is made of the operators of the elliptic and hyperbolic quadrics which are not in the doily \cite{LHS}. 
 \begin{table}
 \begin{center}
\begin{tabular}{c|c|c}
  Geometry & Irreducible representation & Invariant\\
 \hline
 Quadratic cone & $1\oplus 15\oplus 15$ rep of $A_5$ & Pfaffian (for the $15$ of $A_5$)\\
  \hline
 Elliptic Quadric & $27$ irrep of $E_6$ &  Cartan's cubic invariant           \\
 \hline
 Hyperbolic Quadric & $35$ irrep of $A_6$ &  7-order invariant  \\
\end{tabular}
\caption{Correspondence between hyperplanes, representations and invariants in the Veldkamp line $(C_{YYY},H_{YYY},H_{III})$.}\label{inva}
\end{center}
\end{table}
\begin{remark}\rm
 The hyperbolic quadric, i.e. the green part of the Veldkamp line Fig. \ref{veldkamline3qubit}, which corresponds to the weight diagram of the $35$-dimensional 
 irreducible representation of $A_6$, can be further decomposed as $35=15\oplus 20$ for the action of 
 $A_5$. In this decomposition the $15$ of $A_5$ corresponds to the doily as detailed at the beginning of the 
 sectio,n while the other symmetric operators,  which accommodate the 
 diagram of the $20$-dimensional irreducible representation 
 of $A_5$, generate the double-six of Mermin pentagrams \cite{LS}. 
\end{remark}

\begin{remark}\rm
 In \cite{LHS} other finite geometric structures, like extended quadrangle, are revealed 
 in connection with sub-parts of this <<magic>> Veldkamp line.
\end{remark}

\section*{Conclusion}
The two geometric constructions presented in this paper have been known in the mathematics community for quite a long time. The concept of auxiliary varieties (secant, duals) has been 
known since the XIXth century, while the notion of Veldkamp space 
of a point-line geometry goes back to the $80$'s of the XXth century. These geometric constructions have
been shown to be useful in quantum information in the past $15$ years, first
to describe quantum paradoxes such as entanglement and contextuality. These geometric approaches could 
be employed in the future to get  insight into some quantum information protocols \cite{HJN}. 
The fact that representation theory of simple Lie algebras acts as symmetry behind the scene could also lead to interesting findings of how to connect geometrically entanglement and 
contextuality.

\subsection*{Acknowledgments}
The first part of this review paper was presented at the international workshop <<Quantum Physics and Geometry>> organized at Levico Terme in July 2017. I would like to thank
the organizers for inviting me to present an overview on my research 
and to contribute to this UMI Lecture Notes. The research  presented in this review has been done collectively; I would like to warmly thank my co-authors 
Jean-Gabriel Luque, Jean-Yves Thibon, Michel Planat, Metod Saniga, P\'eter L\'evay and Hamza Jaffali 
for our rich collaboration over the past 6 years.
This work was partially supported by the French ``Investissements d'Avenir'' program, project ISITE-BFC (contract ANR-15-IDEX-03).


\begin{thebibliography}{}
\bibitem{amselem} Amselem, E., R\aa{}dmark, M., Bourennane, M., \& Cabello, A. (2009). State-independent quantum contextuality with single photons. Physical Review Letters, 103(16), 160405.
\bibitem{Arkhipov} Arkhipov, A. (2012). Extending and characterizing quantum magic games. arXiv preprint arXiv:1209.3819.
\bibitem{Ar} Arnold, V. I. (1974). Critical points of smooth functions. In Proceedings of ICM-74 (Vol. 1, pp. 19-40).
\bibitem{Ar2} Arnold, V. I. (1981). Singularity theory (Vol. 53). Cambridge University Press.
\bibitem{aspect} Aspect, A., Dalibard, J., \& Roger, G. (1982). Experimental test of Bell's inequalities using time-varying analyzers. Physical Review letters, 49(25), 1804.
\bibitem{Aulbach} Aulbach, M. (2012). Classification of entanglement in symmetric states. International Journal of Quantum Information, 10(07), 1230004.
\bibitem{Aulbach2} Aulbach, M., Markham, D., \& Murao, M. (2010). Geometric Entanglement of Symmetric States and the Majorana Representation. In TQC (pp. 141-158).
\bibitem{bartosik} Bartosik, H., Klepp, J., Schmitzer, C., Sponar, S., Cabello, A., Rauch, H., \& Hasegawa, Y. (2009). Experimental test of quantum contextuality in neutron interferometry. Physical Review Letters, 103(4), 040403.
\bibitem{bell} Bell, J. S. (1966). On the problem of hidden variables in quantum mechanics. Reviews of Modern Physics, 38(3), 447.
\bibitem{bennett} Bennett, C. H., Popescu, S., Rohrlich, D., Smolin, J. A., \& Thapliyal, A. V. (2000). Exact and asymptotic measures of multipartite pure-state entanglement. Physical Review A, 63(1), 012307.
\bibitem{BDD}  Borsten, L., Dahanayake, D., Duff, M. J., Marrani, A., \& Rubens, W. (2010). Four-qubit entanglement classification from string theory. Physical Review Letters, 105(10), 100507.
 \bibitem{BDDER}  Borsten, L., Dahanayake, D., Duff, M. J., Rubens, W., \& Ebrahim, H. (2009). Freudenthal triple classification of three-qubit entanglement. Physical Review A, 80(3), 032326. 	
\bibitem{BDL} Borsten, L., Duff, M. J., \& L\'evay, P. (2012). The black-hole/qubit correspondence: an up-to-date review. Classical and Quantum Gravity, 29(22), 224008.
 \bibitem{BLT1} Briand, E., Luque, J. G., \& Thibon, J. Y. (2003). A complete set of covariants of the four qubit system. Journal of Physics A: Mathematical and General, 36(38), 9915.
 \bibitem{BLT2} Briand, E., Luque, J. G., Thibon, J. Y., \& Verstraete, F. (2004). The moduli space of three-qutrit states. Journal of mathematical physics, 45(12), 4855-4867.
 \bibitem{brody2}  Brody D.C.,  Gustavsson A.C.T. \& Hughston L.P. (2007). Entanglement of three-qubit geometry. J. Phys. Conf. Ser. {\bf 67}, 012044 (2007).
 \bibitem{Bry} Brylinski, J. L. (2002). Algebraic measures of entanglement. Mathematics of quantum computation, 3-23.
 \bibitem{cabello1} Cabello, A. (2008). Experimentally testable state-independent quantum contextuality. Physical Review Letters, 101(21), 210401.

 \bibitem{cabello} Cabello, A., Estebaranz, J., \& Garc\'ia-Alcaine, G. (1996). Bell-Kochen-Specker theorem: A proof with 18 vectors. Physics Letters A, 212(4), 183-187.
\bibitem{CDG} Chen, L., \DJ okovi\'c, D. \v{Z}., Grassl, M., \& Zeng, B. (2013). Four-qubit pure states as fermionic states. Physical Review A, 88(5), 052309.
\bibitem{CD} Chterental, O., \& \DJ okovi\'c, D. \v{Z}. (2006). Normal forms and tensor ranks of pure states of four qubits. arXiv preprint quant-ph/0612184.
\bibitem{chrys} Chryssomalakos, C., Guzman, E., \& Serrano-Ens\'astiga, E. (2017). Geometry of spin coherent states. arXiv preprint arXiv:1710.11326.
\bibitem{cleve} Cleve, R., \& Mittal, R. (2014). Characterization of binary constraint system games. In International Colloquium on Automata, Languages, and Programming (pp. 320-331). Springer, Berlin, Heidelberg.
\bibitem{DF} Duff, M. J., \& Ferrara, S. (2007). $E_7$ and the tripartite entanglement of seven qubits. Physical Review D, 76(2), 025018.
\bibitem{Dur} D\"ur, W., Vidal, G., and Cirac, J. I. (2000). Three qubits can be entangled in two inequivalent ways. Physical Review A, 62(6), 062314.

\bibitem{EPR} Einstein, A., Podolsky, B., \& Rosen, N. (1935). Can quantum-mechanical description of physical reality be considered complete?. Physical Review, 47(10), 777.
\bibitem{FH} Fulton, W.,  Harris, J. (1991). Representation theory (Vol. 129). Springer Science \& Business Media.
\bibitem{GKZ} Gelfand, I. M., Kapranov, M., and Zelevinsky, A. (2008). Discriminants, resultants, and multidimensional determinants. Springer Science $\And$ Business Media
 \bibitem{Ha} Harris, J. (2013). Algebraic geometry: a first course (Vol. 133). Springer Science \& Business Media.
\bibitem{HOS} Havlicek, H., Odehnal, B., \& Saniga, M. (2009). Factor-group-generated polar spaces and (multi-) qudits. Symmetry, Integrability and Geometry. Methods and Applications, 5.
\bibitem{HJ} Holweck, F., \& Jaffali, H. (2016). Three-qutrit entanglement and simple singularities. Journal of Physics A: Mathematical and Theoretical, 49(46), 465301.
\bibitem{HJN} Holweck, F., Jaffali, H., \& Nounouh, I. (2016). Grover's algorithm and the secant varieties. Quantum Information Processing, 15(11), 4391-4413.
\bibitem{HL} Holweck, F., \& L\'evay, P. (2016). Classification of multipartite systems featuring only $\ket{W}$ and $\ket{GHZ}$ genuine entangled states. Journal of Physics A Mathematical General, 49(8).
\bibitem{HLT} Holweck, F., Luque, J. G., $\And$ Thibon, J. Y. (2012). Geometric descriptions of entangled states by auxiliary varieties. Journal of Mathematical Physics, 53(10), 102203.
\bibitem{HLT2} Holweck, F., Luque, J. G., \& Thibon, J. Y. (2014). Entanglement of four qubit systems: A geometric atlas with polynomial compass I (the finite world). Journal of Mathematical Physics, 55(1), 012202.
\bibitem{HLT3} Holweck, F., Luque, J. G., \& Thibon, J. Y. (2017). Entanglement of four-qubit systems: A geometric atlas with polynomial compass II (the tame world). Journal of Mathematical Physics, 58(2), 022201.
\bibitem{HLP} Holweck, F., Luque, J. G., \& Planat, M. (2014). Singularity of type D4 arising from four-qubit systems. Journal of Physics A: Mathematical and Theoretical, 47(13), 135301.
\bibitem{HS} Holweck, F., \& Saniga, M. (2017). Contextuality with a small number of observables. International Journal of Quantum Information, 15(04), 1750026.
\bibitem{HSL} Holweck, F., Saniga, M., \& L\'evay, P. (2014). A Notable Relation between N-Qubit and $2^{N-1}$-Qubit Pauli Groups via Binar $LGr(N, 2N)$. SIGMA. Symmetry, Integrability and Geometry: Methods and Applications, 10, 041.
\bibitem{Hey} Heydari, H. (2008). Geometrical structure of entangled states and the secant variety. Quantum Information Processing, 7(1), 43-50.
\bibitem{HWVE} Howard, M., Wallman, J., Veitch, V., \& Emerson, J. (2014). Contextuality supplies the/magic/'for quantum computation. Nature, 510(7505), 351-355.
\bibitem{KPV}  Kac, V. G., Popov, V. L., Vinberg, E.B.  (1976). Sur les groupes lineaires algebriques dont l'algebres des invariants est libres. CR Acad. Sci. Paris, 283, 875-878.
\bibitem{Kirchmair} Kirchmair, G., Z\"ahringer, F., Gerritsma, R., Kleinmann, M., G\"uhne, O., Cabello, A., Blatt, R. \& Roos, C. F. (2009). State-independent experimental test of quantum contextuality. Nature, 460(7254), 494-497.
\bibitem{KS} Kochen, S., \& Specker, E. P. (1975). The problem of hidden variables in quantum mechanics. In The logico-algebraic approach to quantum mechanics (pp. 293-328). Springer Netherlands.
\bibitem{Landsberg} Landsberg, J. M. (2012) {\em Tensors: geometry and applications}. American Mathematical Society.
\bibitem{LM2} Landsberg, J. M., \& Manivel, L. (2001). The projective geometry of Freudenthal's magic square. Journal of Algebra, 239(2), 477-512.
\bibitem{LM1} Landsberg, J. M., \& Manivel, L. (2004). On the ideals of secant varieties of Segre varieties. Foundations of Computational Mathematics, 4(4), 397-422.
\bibitem{LO} Landsberg, J. M., \& Ottaviani, G. (2013). Equations for secant varieties of Veronese and other varieties. Annali di Matematica Pura ed Applicata, 192(4), 569-606.
\bibitem{LePai}Le Paige C. (1881). Sur la th\'eorie des formes binaires \`a plusieurs s\'eries de  variables. Bull. Acad. Roy. Sci. Belgique  {\bf 2} (3), 40-53.
\bibitem{Levay1} L\'evay, P. (2007). Strings, black holes, the tripartite entanglement of seven qubits, and the Fano plane. Physical Review D, 75(2), 024024.
\bibitem{LH} L\'evay, P., \& Holweck, F. (2015). Embedding qubits into fermionic Fock space: Peculiarities of the four-qubit case. Physical Review D, 91(12), 125029.
\bibitem{LHS} L\'evay, P., Holweck, F., \& Saniga, M. (2017). Magic three-qubit Veldkamp line: A finite geometric underpinning for form theories of gravity and black hole entropy.  Physical Review D 96, 026018
\bibitem{LPS} L\'evay, P., Planat, M., \& Saniga, M. (2013). Grassmannian connection between three-and four-qubit observables, Mermin's contextuality and black holes. Journal of High Energy Physics, 2013(9), 37.
\bibitem{LSVP} L\'evay, P., Saniga, M., Vrana, P., \& Pracna, P. (2009). Black hole entropy and finite geometry. Physical Review D, 79(8), 084036.
\bibitem{LS} L\'evay, P., \& Szab\'o, Z. (2017). Mermin pentagrams arising from Veldkamp lines for three qubits. Journal of Physics A: Mathematical and Theoretical, 50(9), 095201.
\bibitem{LV} L\'evay, P., \& Vrana, P. (2008). Three fermions with six single-particle states can be entangled in two inequivalent ways. Physical Review A, 78(2), 022329.
\bibitem{LT1} Luque, J. G., \& Thibon, J. Y. (2003). Polynomial invariants of four qubits. Physical Review A, 67(4), 042303.
\bibitem{LT2} Luque, J. G., \& Thibon, J. Y. (2005). Algebraic invariants of five qubits. Journal of Physics A: Mathematical and General, 39(2), 371.
\bibitem{mermin} Mermin, N. D. (1993). Hidden variables and the two theorems of John Bell. Reviews of Modern Physics, 65(3), 803.
\bibitem{My} Miyake, A. (2003). Classification of multipartite entangled states by multidimensional determinants. Physical Review A, 67(1), 012108.
\bibitem{My2} Miyake, A., \& Verstraete, F. (2004). Multipartite entanglement in 2x2xn quantum systems. Physical Review A, 69(1), 012101.
\bibitem{My3} Miyake, A., \& Wadati, M. (2002). Multipartite entanglement and hyperdeterminants. Quantum Information \& Computation, 2(7), 540-555.
\bibitem{Nu} Nurmiev, A. G. (2000). Orbits and invariants of third-order matrices. Mat. Sb., 191(5), 101-108.
\bibitem{O1} Oeding, L. (2011). Set-theoretic defining equations of the tangential variety of the Segre variety. Journal of Pure and Applied Algebra, 215(6), 1516-1527.
\bibitem{Pav} Parfenov, P. G. (1998). Tensor products with finitely many orbits. Russian Mathematical Surveys, 53(3), 635-636.
\bibitem{PS} Planat, M., \& Saniga, M. (2012). Five-qubit contextuality, noise-like distribution of distances between maximal bases and finite geometry. Physics Letters A, 376(46), 3485-3490.
\bibitem{PSH} Planat, M., Saniga, M., \& Holweck, F. (2013). Distinguished three-qubit <<magicity>> via automorphisms of the split Cayley hexagon. Quantum Information Processing, 12(7), 2535-2549.
\bibitem{peres} Peres, A. (1991). Two simple proofs of the Kochen-Specker theorem. Journal of Physics A: Mathematical and General, 24(4), L175.
\bibitem{SHHPP} Saniga, M., Havlicek, H., Holweck, F., Planat, M., \& Pracna, P. (2015). Veldkamp-space aspects of a sequence of nested binary Segre varieties. Annales de l'Institut 
Henri Poincar\'e D, 2(3), 309-333.
\bibitem{SPPH} Saniga, M., Planat, M., Pracna, P., \& Havlicek, H. (2007). The Veldkamp space of two-qubits. Symmetry, Integrability and Geometry. Methods and Applications, 3.
\bibitem{SP1} Saniga, M., \& Planat, M. (2007). Multiple Qubits as Symplectic Polar Spaces of Order Two. Advanced Studies in Theoretical Physics, 1, 1-4.
\bibitem{SP} Saniga, M., \& Planat, M. (2012). Finite geometry behind the Harvey-Chryssanthacopoulos 
four-qubit magic rectangle. Quantum Information \& Computation, 12(11-12), 1011-1016.
\bibitem{sanz} Sanz, M., Braak, D., Solano, E., \& Egusquiza, I. L. (2017). Entanglement classification with algebraic geometry. Journal of Physics A: Mathematical and Theoretical, 50(19), 195303.
\bibitem{Sarosi} S\'arosi, G., \& L\'evay, P. (2014). Entanglement in fermionic Fock space. Journal of Physics A: Mathematical and Theoretical, 47(11), 115304.
\bibitem{Sawicki} Sawicki, A., \& Tsanov, V. V. (2013). A link between quantum entanglement, secant varieties and sphericity. Journal of Physics A: Mathematical and Theoretical, 46(26), 265301.
\bibitem{Thas} Thas, K. (2009). The geometry of generalized Pauli operators of N-qudit Hilbert space, and an application to MUBs. EPL (Europhysics Letters), 86(6), 60005.
\bibitem{Terra} Terracini, A. (1911). Sulle $v_k$ per cui la varieta degli $s_h(h+ 1)$ seganti ha dimensione minore dell'ordinario. Rendiconti del Circolo Matematico di Palermo (1884-1940), 31(1), 392-396.
\bibitem{V} Verstraete, F., Dehaene, J., De Moor, B., \& Verschelde, H. (2002). Four qubits can be entangled in nine different ways. Physical Review A, 65(5), 052112.
\bibitem{VL2} Vrana, P., \& L\'evay, P. (2009). Special entangled quantum systems and the Freudenthal construction. Journal of Physics A: Mathematical and Theoretical, 42(28), 285303.
\bibitem{VL} Vrana, P., \& L\'evay, P. (2010). The Veldkamp space of multiple qubits. Journal of Physics A: Mathematical and Theoretical, 43(12), 125303.
\bibitem{WA1} Waegell, M., \& Aravind, P. K. (2012). Proofs of the Kochen-Specker theorem based on a system of three qubits. Journal of Physics A: Mathematical and Theoretical, 45(40), 405301.
\bibitem{WA2} Waegell, M., \& Aravind, P. K. (2013). Proofs of the Kochen-Specker theorem based on the N-qubit Pauli group. Physical Review A, 88(1), 012102.
\bibitem{WZ} Weyman, J., \& Zelevinsky, A. (1996). Singularities of hyperdeterminants. In Annales de l'Institut Fourier (Vol. 46, No. 3, pp. 591-644). 
\bibitem{Zak} Zak, F. L. (1993). Tangents and secants of algebraic varieties, Translations of Mathematical Monographs, vol. 127. American Mathematical Society, Providence, RI.
\end{thebibliography}
\end{document}